\documentclass[aps,prd,reprint,amsmath,amssymb,floatfix]{revtex4-1}

\pdfoutput=1

\usepackage{epsfig}
\usepackage{amssymb,amsmath}
\usepackage{epstopdf}
\usepackage{url}

\def\be{\begin{equation}}
\def\ee{\end{equation}}
\def\bea{\begin{eqnarray}}
\def\eea{\end{eqnarray}}
\def\beal{\begin{align}}
\def\eeal{\end{align}}

\newcommand{\Kahler}{\textrm{K\"{a}hler}~}

\relax
\begin{document}

\title{BMSSM Higgs Bosons at the Tevatron and the LHC}


\author{Marcela Carena}
\affiliation{Theoretical Physics Department, Fermilab, Batavia, Illinois 60510, USA}
\affiliation{Enrico Fermi Institute, University of Chicago, 5640 Ellis Avenue, Chicago, Illinois 60637, USA}
\author{ Eduardo Pont\'{o}n}
\affiliation{ Department of Physics, Columbia University,
538 W. 120th St., New York, New York 10027, USA}
\author{ Jos\'{e} Zurita}
\affiliation{Institut f{\"u}r Theoretische Physik, Universit\"at Z\"urich, Winterthurerstrasse 190, CH-8057 Z\"urich, Switzerland. }

\begin{abstract}
We study extensions of the minimal supersymmetric standard model
(MSSM) with new degrees of freedom that couple sizably to the MSSM
Higgs sector and lie in the TeV range.  After integrating out the
physics at the TeV scale, the resulting Higgs spectrum can
significantly differ from typical supersymmetric scenarios, thereby
providing a window beyond the MSSM (BMSSM).  Taking into account
current LEP and Tevatron constraints, we perform an in-depth analysis
of the Higgs collider phenomenology and explore distinctive
characteristics of our scenario with respect to both the standard
model and the MSSM. We propose benchmark scenarios to illustrate
specific features of BMSSM Higgs searches at the Tevatron and the LHC.
\end{abstract}


\maketitle


\section{Introduction}
\label{sec:intro}
There has been a recent surge of interest in extensions of the minimal
supersymmetric standard model (MSSM) by higher-dimension
operators~\cite{Strumia:1999jm,Brignole:2003cm,Dine:2007xi,Antoniadis:2007xc,Antoniadis:2009rn,Randall:2007as,Batra:2008rc,Carena:2009gx,Blum:2009na,Casas:2003jx,Cassel:2009ps}.
These can have an important impact on the Higgs sector, alleviating in
particular the tension present in the MSSM that results from the LEP
Higgs bounds.  Such effective field theory (EFT) studies allow a
model-independent description of a large class of extensions of the
MSSM, and permit one to quantify the sense in which the Higgs sector
can serve as a window beyond the MSSM (BMSSM).

This point of view was clearly put forward in Ref.~\cite{Dine:2007xi},
where it was emphasized that at leading order in $1/M$ --where $M$ is
the scale of the physics that is integrated out-- the MSSM is extended
by only two parameters.  The surprisingly large effects of such
higher-dimension operators can be understood from the fact that the
MSSM Higgs potential is rather restricted at tree-level.  The
nonrenormalizable operators in the superpotential induce
renormalizable (quartic) operators in the Higgs potential that are not
present in the MSSM limit (at tree-level), so that in spite of the
fact that their coefficients are ``small'' --of order $\mu/M$-- they
correspond to qualitatively new effects.  In fact, the operators thus
induced can easily ``destabilize'' the MSSM-like minimum and lead to
new minima that exist only as a direct result of the higher-dimension
operators (i.e. the heavy physics).  It was emphasized in
\cite{Batra:2008rc} that such minima can be phenomenologically viable,
can be studied within the EFT framework, and can explain the distinct
properties induced by the heavy physics on the Higgs sector.

If the BMSSM physics is sufficiently heavy, the leading order analysis
at order $1/M$ can suffice. However, it is perfectly possible that
$M$ is not too far from the electroweak (EW) scale, and that
nevertheless the heavy physics may not be easy to see directly at the
LHC, even if it is within its kinematic reach (e.g. heavy singlets
that couple only through the Higgs sector).  In such cases, the EFT
approach is still useful to describe the properties of the MSSM Higgs
sector.  It turns out that the effects of order $1/M^{2}$ are more
important than naively expected.  This observation also finds a simple
explanation in the structure of the MSSM tree-level Higgs
potential~\cite{Carena:2009gx} together with the smallness of the MSSM
tree-level quartic couplings (the root cause for a Higgs state lighter
than the $Z$ mass in the MSSM at tree-level).  The crucial point is
that the leading order contributions to a subset of the quartic Higgs
operators, from the heavy physics, first enter at order $1/M^{2}$.
Thus, these end up correcting a coefficient of order $g^{2}$ instead
of a $1/M$ effect, and can give a relevant correction even if the
expansion parameter is relatively small.  Nevertheless, it is
important to appreciate that the fact that the first two orders in the
expansion in $1/M$ can even result in comparable contributions to the
Higgs masses, in no way implies a breakdown of the EFT.

In Ref.~\cite{Carena:2009gx} a detailed study of the consequences for
the Higgs masses and couplings up to order $1/M^{2}$ was given, and a
selected number of phenomenological observations were already made,
such as: enhanced gluon fusion production cross sections in a large
number of cases, and the presence of ``exotic'' decay modes with more
than one Higgs boson in the decay chain.  In this work we analyze the
constraints from LEP and the Tevatron on the neutral Higgs bosons, as
well as the charged Higgs bounds from LEP~\footnote{The present
Tevatron bounds on supersymmetric charged Higgs bosons are beyond the
parameter space that we study in this work.}.  We also expand on the
associated collider phenomenology, emphasizing the type of signals
that can be expected at both the Tevatron and the LHC. We point out
that due to the large corrections to the Higgs masses (especially to
the CP-even Higgs bosons) the production and decay patterns can be
markedly different from those in the MSSM. Examples include models
where both CP even Higgs bosons have significant branching fractions
into gauge bosons, thus giving rise to spectacular signals such as two
clearly defined peaks in the di-lepton invariant mass distribution.
In addition, we observe new decay chains that allow for production of
the ``nonstandard'' Higgs bosons without large $\tan\beta$
enhancements.  It is possible that the full two-Higgs-doublet-model
(2HDM) content can be mapped in detail, thus providing a clear and
definite signal for physics beyond the standard model, and a rather
detailed understanding of the mechanism of electroweak symmetry
breaking (EWSB).  If, in addition, relatively light superparticle
signals are observed, as might be expected in these scenarios, a clear
case for BMSSM physics could be established.  Apart from the collider
phenomenology induced indirectly by the heavy physics,
higher-dimension operators have also been studied in the context of
dark matter~\cite{Cheung:2009qk,Berg:2009mq,Bernal:2009jc},
cosmology~\cite{Bernal:2009hd} and EW
baryogenesis~\cite{Grojean:2004xa,Bodeker:2004ws,Delaunay:2007wb,Noble:2007kk,Blum:2008ym},
and it may be interesting to further explore the connections with
collider physics.

This paper is organized as follows.  In Sec.~\ref{sec:review}, we
summarize the most relevant aspects of the models under study.  In
Sec.~\ref{sec:spectra} we discuss the modifications of the Higgs
spectrum, which are the dominant factor in determining the Higgs
collider phenomenology.  In Sec.~\ref{sec:results} we discuss in
detail the range of signatures uncovered by our survey, separating the
analysis into the low and large $\tan\beta$ regimes.  We conclude in
Sec.~\ref{sec:conclu}.

\section{Extended SUSY Higgs Sectors at a Glance}
\label{sec:review}

As already mentioned, when considering BMSSM scenarios where the
non-MSSM degrees of freedom have masses parametrically larger than the
weak scale, an EFT approach is very useful.  The fact that at leading
order only two parameters are added to those in the MSSM makes this a
rather economic extension~\cite{Dine:2007xi}, that nevertheless can
significantly change the MSSM Higgs phenomenology.  However, the same
reason that makes these $1/M$ suppressed effects rather important also
implies that the next order in the $1/M$ expansion can be
phenomenologically relevant, without implying a breakdown of the
EFT~\cite{Carena:2009gx}.  At order $1/M^{2}$ there are several
SUSY-preserving and SUSY-violating operators in the \Kahler potential,
the most important of which, in relation to the Higgs phenomenology,
were listed in Ref.~\cite{Carena:2009gx}.  We refer the reader to this
reference for the detailed form of such operators and how they affect
the expressions for the Higgs masses and couplings.  Here we restrict ourselves
to a few general remarks that summarize the most relevant features for
the present study (full details were given in the above reference).

First, it has to be pointed out that the higher-dimension operators to
order $1/M^{2}$ can be easily generated from UV completions that
include a combination of Higgs singlets, $SU(2)$ Higgs triplets, heavy
W primes and Z primes.  As argued in \cite{Carena:2009gx} the upshot
is that the coefficients of the higher-dimension operators, from a
low-energy point of view, can be chosen in an uncorrelated manner.
Although the EFT description to order $1/M^{2}$ introduces a large
number of parameters, which makes the framework more involved compared
to the truncation at order $1/M$, one should notice that this same
feature gives additional handles to infer properties of the heavy
sector from the properties of the low-energy degrees of freedom.  In
any case, since our goal is to survey the collider signal
possibilities in a model-independent way (in a supersymmetric
framework), we focus on a low-energy study based on the EFT at order
$1/M^{2}$, as described in~\cite{Carena:2009gx}.~\footnote{We have
checked that these results are consistent, in the appropriate limits,
with those of Ref.~\cite{Antoniadis:2009rn}, that appeared soon after
Ref.~\cite{Carena:2009gx}.  We emphasize, however, that one has to
treat near degenerate cases in the CP-even sector with care, as
explained in \cite{Carena:2009gx}.}

A random scan over parameter space was performed, and a set of points
satisfying several constraints was selected.  The set of points in
this study satisfy:
\begin{itemize}

\item All the dimensionless coefficients parametrizing the
higher-dimension operators are taken to be at most of order one, i.e.
it is assumed that the heavy physics at $M$ is weakly coupled.

\item \textit{Global minimum}: since the scalar potential can present
several minima, we make sure that the vacuum under study is the global
one (at least within the EFT).  We also check that there are no
charge/color breaking minima, and for simplicity we restrict to the CP
conserving case (checking that the global minimum does not break CP
spontaneously).

\item \textit{Robustness}: there are no accidental cancellations that
can render (not computed) higher orders in the $1/M$ expansion more
important than expected.

\item \textit{``Light'' SUSY spectrum}: given that generically, and
unlike in the MSSM, these models satisfy the LEP bounds on the Higgs
mass at tree-level, there is no need for large radiative corrections.
Naturalness suggests that in these models the SUSY spectrum would be
expected to be light (in the few hundred GeV range, consistent with
direct bounds).

\item Agreement with EW precision constraints, in particular in
regards to the Peskin-Takeuchi $T$ parameter~\cite{Peskin:1990zt}.
These arise from three sources: a subset of the higher-dimension
operators (as generated, for instance, by Higgs triplets), the details
of the MSSM Higgs spectrum, and potential custodially-violating mass
splittings in the sparticle spectrum.  We emphasize that mild
cancellations allow for higher-dimension operator effects that can
have a non-negligible impact on the Higgs collider phenomenology.

\end{itemize}

All of the above constraints were described in detail in
\cite{Carena:2009gx}.  In addition, we impose the current collider
bounds from LEP and the Tevatron using the code HiggsBounds v1.2.0
\cite{Bechtle:2008jh,Bechtle:2009ic}.~\footnote{We thank the authors
of \cite{Bechtle:2008jh,Bechtle:2009ic} for providing us with a
modified version of the code that includes the LEP 2 jet analysis.} To
this we add the LEP bounds on charged Higgs production \cite{:2001xy},
and the newest combined result from the Tevatron in the WW channel
\cite{Aaltonen:2010yv}, and in the inclusive tau search
\cite{:tau_inclusive}, that are not included in the currently
available version of this code.  We use HiggsBounds in the ``effective
coupling'' mode, which requires effective couplings defined by
\bea
g_{\phi X}^2 &=& \frac{\Gamma(\phi \to X)}{\Gamma_{SM}(\phi \to X)}~,
\label{geff}
\eea
where $\phi = h, H, A$ is any of the neutral Higgs states,
$\Gamma(\phi \to X)$ is the partial width in our model into any of the
final states $X =
s\bar{s},c\bar{c},b\bar{b},\tau\bar{\tau},WW,ZZ,\gamma \gamma$ or $g
g$ (when applicable), and $\Gamma_{SM}(\phi \to X)$ is the partial
width for a SM Higgs of the corresponding mass.  Together with the
total widths in our model (and in the SM), these effective couplings
encode the information about branching fractions into these decay
channels in our model.

We have implemented our tree-level expressions for the spectrum and
Higgs couplings in HDECAY v3.4~\cite{Djouadi:1997yw}.  This allows us
to compute the Higgs partial decay widths, taking into account the QCD
radiative corrections, that are known to be sizable (for a review, see
\cite{Spira:1997dg}).  In addition, we include the radiative
corrections derived from the 1-loop RG improved effective potential
due to supersymmetric particles~\cite{Carena:1995bx}, and the SUSY
QCD/EW corrections to the Yukawa
couplings~\cite{Carena:1994bv,Carena:1999py}.  Loop contributions from
the heavy physics that has been integrated out are suppressed by both
a loop factor and by powers of $M$, hence they are expected to be
negligible.

In all the plots that follow, we have fixed the following dimensionful
parameters: $M=1~{\rm TeV}$, $\mu = m_{S} = 200~{\rm
GeV}$,~\footnote{Here $m_{S}$ gives the scale of SUSY breaking in the
heavy sector.  The detailed differences between the various
SUSY-breaking operators are parametrized via ${\cal O}(1)$
dimensionless parameters over which we scan.  See \cite{Carena:2009gx}
for complete details.} and for simplicity, we use a common value
$M_{SUSY} = 300~{\rm GeV}$ and $A_{t} = A_{b} = 0$ in the MSSM
sparticle sector.\footnote{We evaluate the scale inside the logarithms
associated with SUSY loops at $\sqrt{M^2_{SUSY} + m^2_{t}} \approx
347~{\rm GeV}$.} The light superparticle spectrum implies that the
loop contributions to the Higgs masses are modest, while the loop
contributions to the Higgs couplings are more important and sensitive
to the details of this
spectrum~\cite{Carena:1994bv,Carena:1999py,Dawson:1996xz}.  The above
choice of $M_{SUSY} = 300~{\rm GeV}$ is simply meant to illustrate the
possible loop effects arising from relatively light superparticles.
In particular, one can expect the first two generation squarks to be
somewhat heavier to satisfy direct collider
bounds~\cite{:2007ww,Aaltonen:2008rv} or the sleptons could be
somewhat lighter, without changing our generic conclusions regarding
the Higgs collider phenomenology.  Note also that the
neutralino/chargino sector depends on parameters not affecting the
Higgs sector directly, and in particular that we do not impose
constraints from dark matter (in this work, we remain agnostic as to
the identity of the DM candidate, but see~\cite{CMR}).  We have also
not imposed indirect constraints, such as those arising from $b
\rightarrow s \gamma$, $B_{s} \to \mu^+ \mu^{-}$ and $g_{\mu} - 2$,
that have the potential to put important restrictions, but depend on
the flavor structure of the soft SUSY-breaking parameters.

We consider two representative values of $\tan\beta$: $\tan\beta = 2$
and $\tan\beta = 20$.  The CP-odd mass was varied in the range
$20-400~{\rm GeV}$.  The upper bound is taken to ensure a proper
separation between the light and heavy scales, as required by the EFT
analysis.  The very low mass range is expected to be severely
constrained, but we defer the study of such region to future work.  We
turn next to a detailed description of the most important physical
characteristics of the set of models in the scan, starting with the
Higgs spectrum.

\section{Masses  of Low-energy Higgs Bosons}
\label{sec:spectra}

In this section we study the spectra of these models, analyzing the
modifications with respect to the MSSM. Compared to the results
already presented in \cite{Carena:2009gx}, we include the 1-loop
supersymmetric corrections to the Higgs quartic couplings as given in
\cite{Carena:1995bx} (a minor effect for the relatively low SUSY
spectrum we have in mind), as well as the constraints coming from
collider data (LEP and Tevatron).

\begin{figure*}[!htp]
\begin{center}
\includegraphics[width=8.0cm]{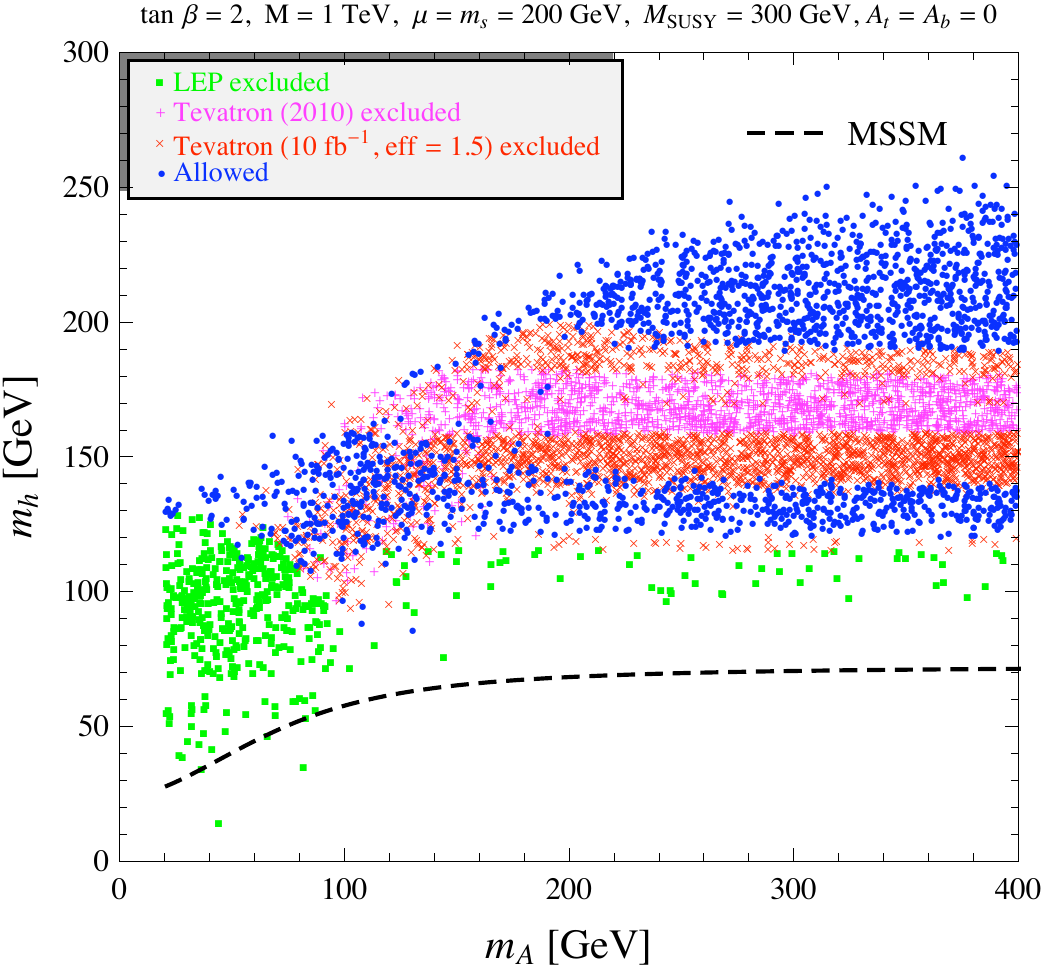}
\hspace{2mm}
\includegraphics[width=8.0cm]{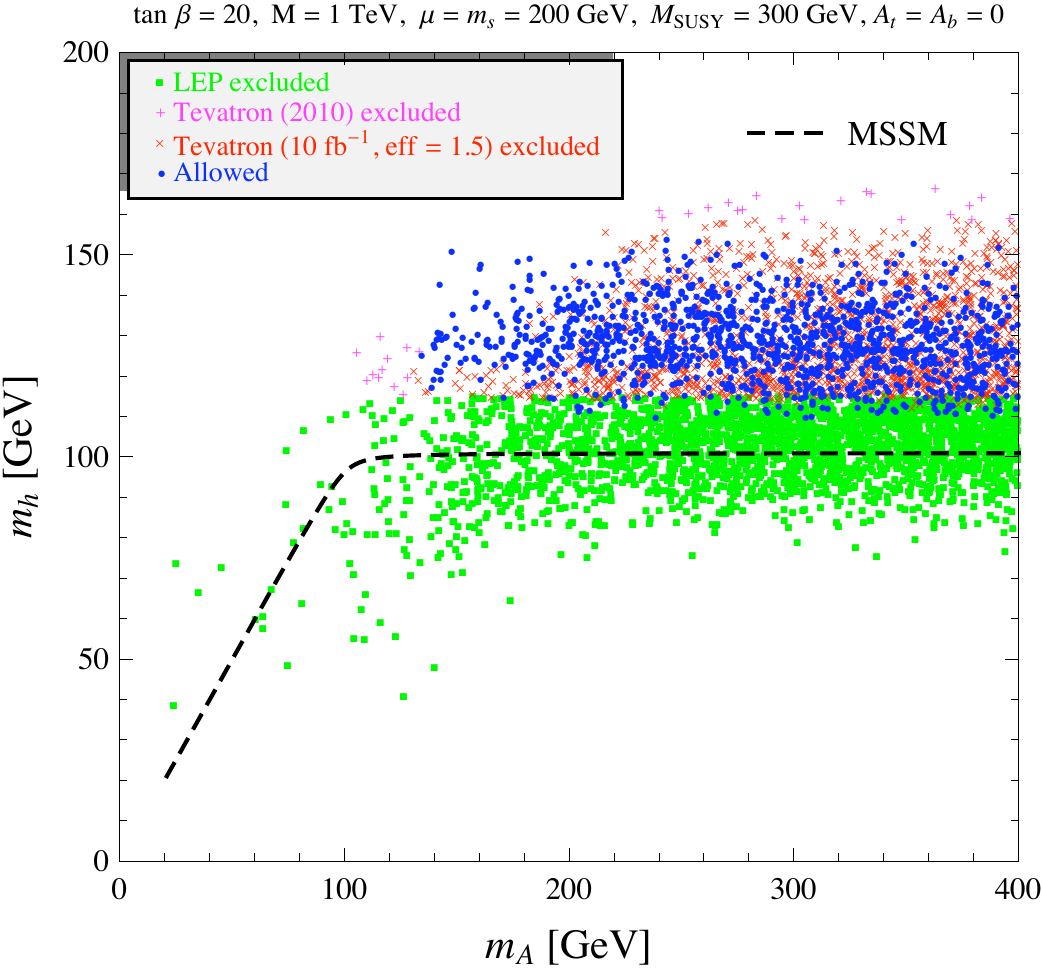}
\end{center}
\vspace{-0.5cm} 
\caption{\label{fig:mlvsma}{\em Lightest CP-even Higgs
boson mass as a function of $m_A$, for $\tan \beta=2$ (left panel) and
$\tan \beta=20$ (right panel).  We show the points excluded by LEP
(green), excluded by current Tevatron data (magenta) and the region
that will be probed by the Tevatron in the near future (red).  The
blue points are allowed by all the current experimental constraints.
The dashed line is the MSSM result for the given SUSY spectrum.  }}
\end{figure*}
In Fig.~\ref{fig:mlvsma} we show the mass of the lightest CP-even
Higgs ($h$) as a function of $m_A$, for both $\tan \beta=2$ (left
panel) and $\tan \beta=20$ (right panel).  The green points represent
models ruled out by LEP, while the magenta ones are excluded by
current data from CDF and D0.  We divide the remaining allowed models
into two subsets.  First, one has those models that will be probed at
the Tevatron at 95 \% C.L (red points), assuming $10~{\rm fb}^{-1}$
per experiment and $50\%$ efficiency
improvements~\cite{Moriond_Fisher}~(see \cite{Draper:2009fh} for
detailed projections in the MSSM context).  These comprise the future
reach of two search channels: $h/H \to b \bar{b}$ with the Higgs being
produced in association with electroweak gauge bosons, and $gg \to h/H
\to W^+ W^-$.  Second, the blue points are those that will be out of
the Tevatron reach under the previous assumptions.  For reference, we
also show the MSSM (dashed) curve, assuming the same light SUSY
spectrum.  This color code will be employed in all the plots.

The corrections to $m_h$ due to the new physics are most important in
the low $\tan \beta$ regime.  Nevertheless, it is clear that they can
also be relevant at large $\tan \beta$.  The higher-dimension
operators affect $m_h$ in such a way that it can easily be above the
MSSM value.  In the left plot, where $\tan \beta=2$, all the points
lie above the MSSM curve; $m_h$ can reach values as high as $250~{\rm
GeV}$.

Moreover, the left panel of Fig.~\ref{fig:mlvsma} shows in a clear way
how the Tevatron probes these models.  For high enough values of $m_A
$ one distinguishes mostly uniformly single colored horizontal
stripes.  The magenta one, where $m_h \sim160-170~{\rm GeV}$,
corresponds to $h$ being excluded by the current Tevatron search in
the $WW$ channel~\cite{Aaltonen:2010yv}.  Note that this range is
slightly larger than the SM one ($162 - 166~{\rm GeV}$).  This is due
to the fact that, in our models, the gluon fusion cross section can be
mildly enhanced with respect to the SM one.  By the same token, one
understands the presence of a few red points within the magenta stripe
as those corresponding to models whose gluon fusion cross section is
below the SM value.  The two red stripes ($m_{h}$ in the ranges $140 -
160~{\rm GeV}$ and $180 - 190~{\rm GeV}$) represent the future
Tevatron reach of the $h \to W^+ W^-$ channel.  Notice also the
presence of a thin stripe of red points, with $m_h$ around $120~{\rm
GeV}$, that extends along a wide range of $m_A$: these models can be
probed by the $h \to b\bar{b} $ channel, that is effective only for
relatively low values of $m_h$.  No points are excluded by the $H \to
b \bar{b}$ decay mode, since $H$ is always much heavier than $120~{\rm
GeV}$.

The two blue stripes correspond to points where there is no reach from
the Tevatron in the $WW$ channel.  This can be explained either by a
low signal due to the reduced branching fraction into gauge bosons, or
simply because the $gg$ parton luminosity is not enough to produce
such a heavy Higgs boson.  Note however that in the high $m_{h}$ blue
region the $ZZ \to 4 l$ channel becomes kinematically accessible, so
that this Higgs could be observed in the \emph{gold plated}
four-lepton mode at the LHC. We will postpone further comments on this
region to the next section.

Regarding the LEP constraints, one sees that there are a few currently
allowed (blue and red) points below the SM LEP bound of $114.4~{\rm
GeV}$\cite{:2001xx,Schael:2006cr}.  The nonexclusion is due to the
fact that the coupling of $h$ to the gauge bosons is reduced with
respect to the SM value.  However, all the red points below the
LEP-bound can potentially be excluded in the $H \to WW$ channel.

For the remaining points ($m_A < 160~{\rm GeV}$, $114.4 ~{\rm GeV}
\lesssim m_h \lesssim 170~{\rm GeV}$), the situation is more complex,
and magenta, blue and red points
coexist in this region.  In particular, there is a region of
allowed (blue) points with $m_h \sim130~{\rm GeV} - 140~{\rm GeV}$ and
relatively low $m_A$.  These points have suppressed branching
fractions into both $WW$ and $b\bar{b}$, with $AA$ being the dominant
decay channel.

In the case of $\tan \beta=20$, the deviations from the MSSM are far
less dramatic.  Ultimately, this is explained by the fact that several
higher-dimension operators are $\tan \beta$ suppressed.  However,
$m_h$ can reach values as high as $160~ {\rm GeV}$.  In this case,
since $h$ is SM-like, the LEP bound is very strict, forcing $m_h$ to
be above $\sim 110~{\rm GeV}$.  Regarding the Tevatron searches, we
see that there are two small and disjoint currently excluded (magenta)
regions.  The region with $m_h$ around $160~{\rm GeV}$ corresponds, as
in the low $\tan\beta$ case, to exclusion based on the $h \to WW$
decay mode.  The second magenta region has lower values of
$m_h~(114-130~{\rm GeV}$) and $m_A ~(100-135 ~{\rm GeV}$).  This
latter set of models are currently excluded by the inclusive tau
search with 2.2 fb$^{-1}$, using the combination from CDF and
D0~\cite{:tau_inclusive}.  This channel becomes important here, since
the $H/A$ --and in some cases the $h$-- coupling to down-type fermions
is $\tan \beta$ enhanced.\footnote{We have not included the future
projection of this channel in our analysis.  We expect that the
increase in luminosity has a minor incidence in the additional number
of points excluded.} Turning to the red points (i.e. those within
future Tevatron sensitivity), a closer inspection reveals that all of
them can be excluded due to the decay modes of the lightest Higgs.  In
more detail, the $h \to b\bar{b}$ channel probes points with $m_{h}$
below $126~{\rm GeV}$, while the rest are probed by the $h \to WW$
search.  This can be understood from the fact that, as in the MSSM, in
the large $\tan \beta$ limit $H$ tends to be non SM-like.  In contrast
to the low $\tan \beta$ case, all the Tevatron allowed (blue) points
correspond to somewhat heavy values of $m_A$ (above $140~{\rm GeV}$).

\begin{figure*}[htp]
\begin{center}
\includegraphics[width=7.9cm]{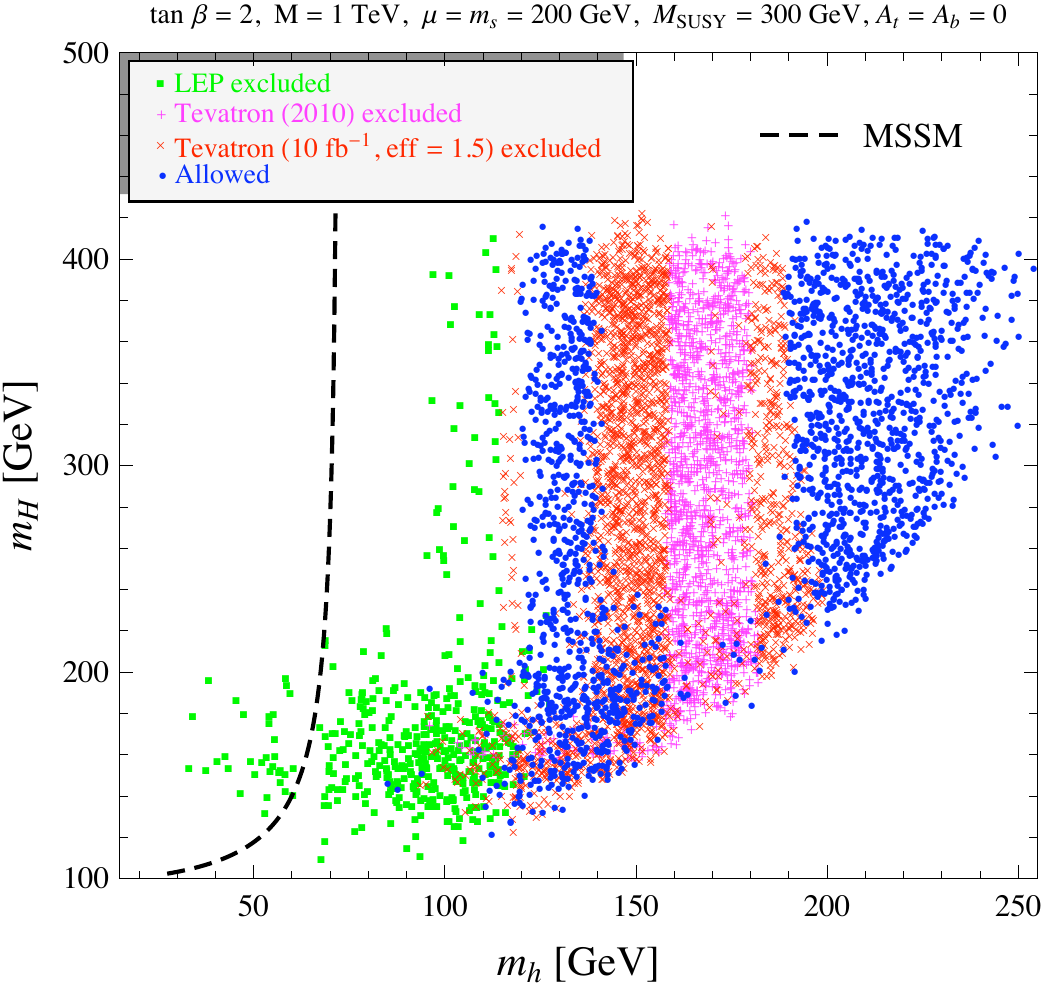}
\hspace{3mm}
\includegraphics[width=7.9cm]{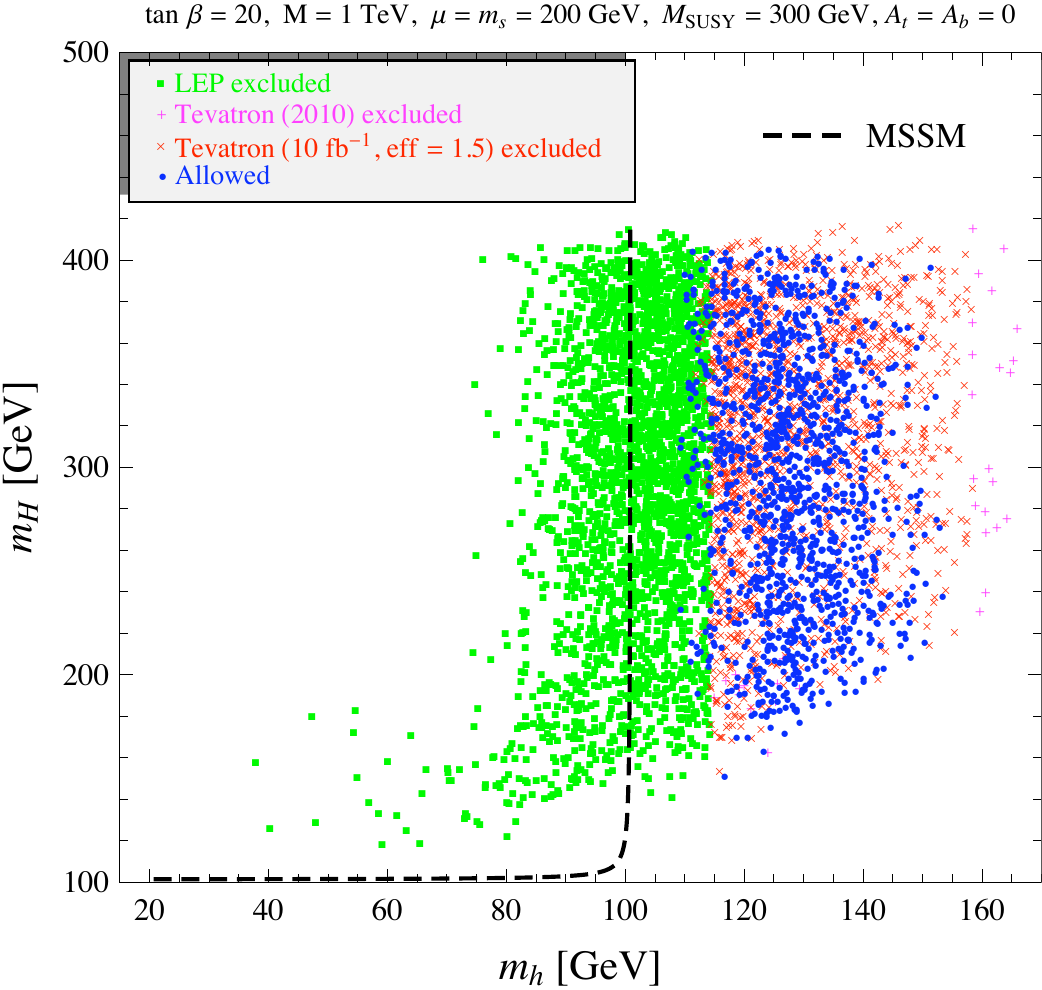}
\end{center}
\vspace{-0.5cm} 
\caption{\label{fig:mhvsml}{\em $m_H$ as a function of
$m_h$, for $\tan \beta=2$ (left panel) and $\tan \beta=20$ (right
panel).  We show the points excluded by LEP (green), current Tevatron
data (magenta) and the region that will be proved by the Tevatron in
the near future (red).  The blue points are allowed by all the current
experimental constraints.  The dashed line is the MSSM result for the
given SUSY spectrum.  }}
\end{figure*}

It is also interesting to study the relation between the CP-even Higgs
masses.  In Fig.~\ref{fig:mhvsml} we show $m_H$ as a function of
$m_h$, for $\tan \beta=2$ (left panel) and $\tan \beta=20$ (right
panel).  For most of the points these masses are not correlated.  For
instance, if in the left plot one takes $m_h$ in the $120-200~{\rm
GeV}$ range, then $M_H$ can vary between $200$ and $400~{\rm GeV}$.
For $\tan \beta=2$, one has not only the (now vertical) stripes
corresponding to exclusion due to $h$ that we have found in
Fig.~\ref{fig:mlvsma}: there are also horizontal stripes,
corresponding to $m_H$ ranges where the Tevatron is excluding models
by means of the $H \to WW$ decay channel.  This sheds some light into
the region already mentioned in the discussion of
Fig.~\ref{fig:mlvsma} with $m_A < 160~{\rm GeV}$ and $114.4 ~{\rm GeV}
\lesssim m_h \lesssim 170~{\rm GeV}$.  In this region both $h$ and $H$
can couple to $WW$, typically resulting in some suppression with
respect to the SM for one or the other CP-even Higgs boson.  This
constitutes an interesting example of how the $h$ and $H$ signals can
complement each other.  The right panel confirms what we have
anticipated from our discussion of Fig.~\ref{fig:mlvsma}: in the large
$\tan \beta$ regime, $m_H$ tends to be heavy, and the decays of $H$
are less restrictive than the ones from $h$, hence there are no
horizontal stripes in this plot.  As mentioned before, the $h$ search
channels give rise to all the red points.

\begin{figure*}[!thp]
\begin{center}
\includegraphics[width=7.9cm]{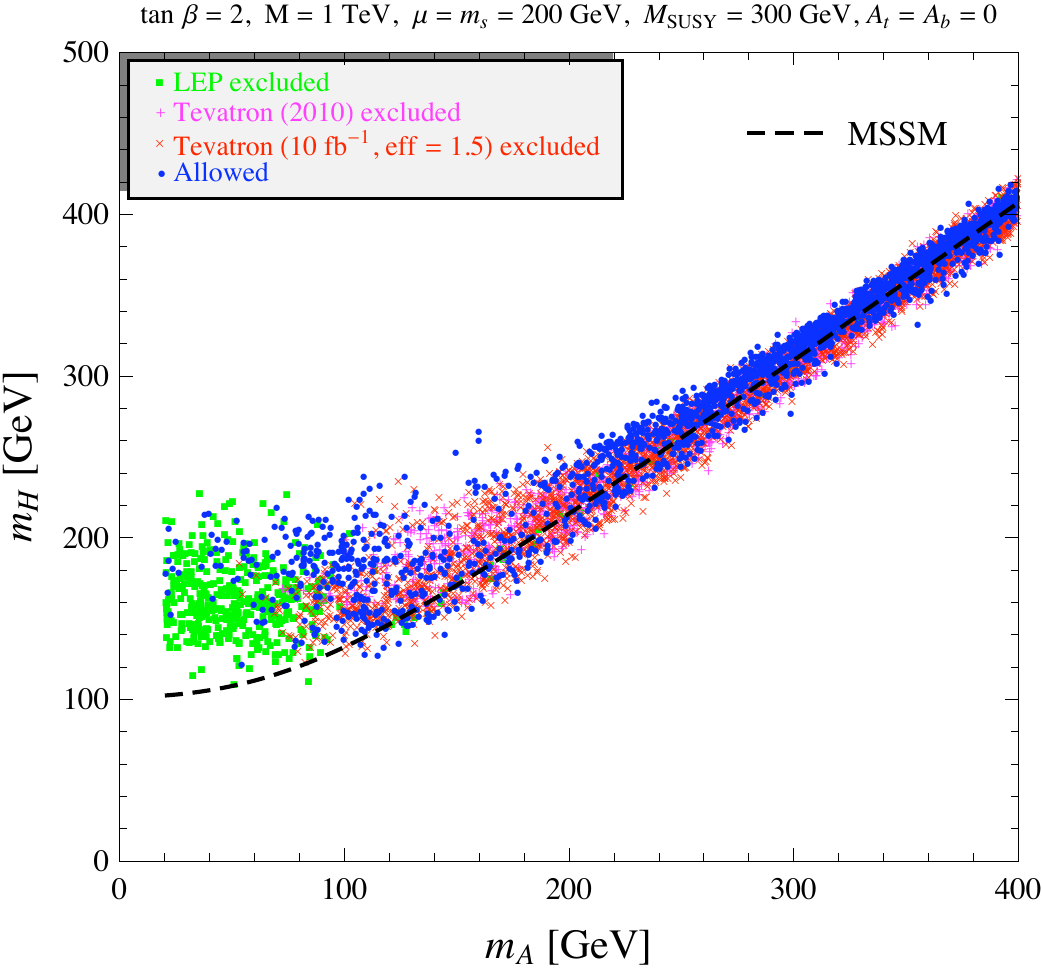}
\vspace{5mm}
\hspace{3mm}
\includegraphics[width =7.9cm]{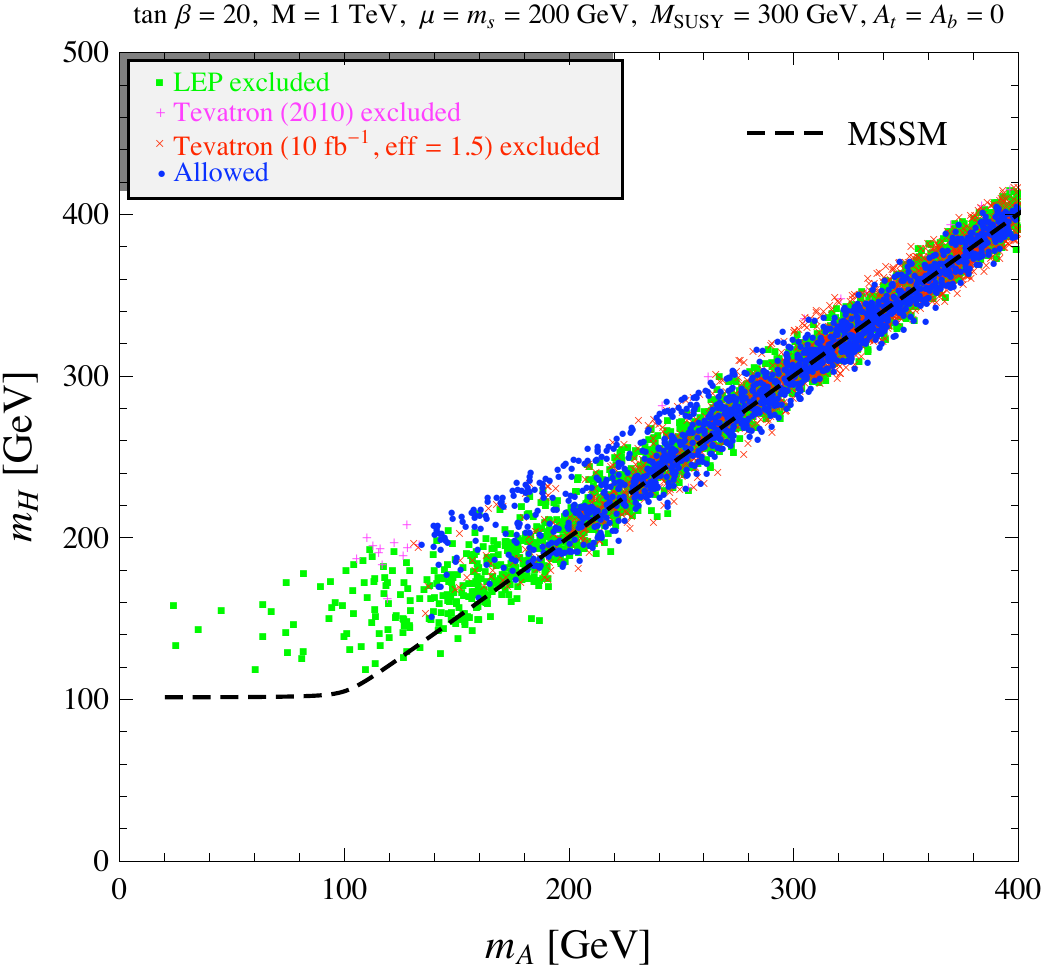}
\vspace{0cm}
\includegraphics[width=7.9cm]{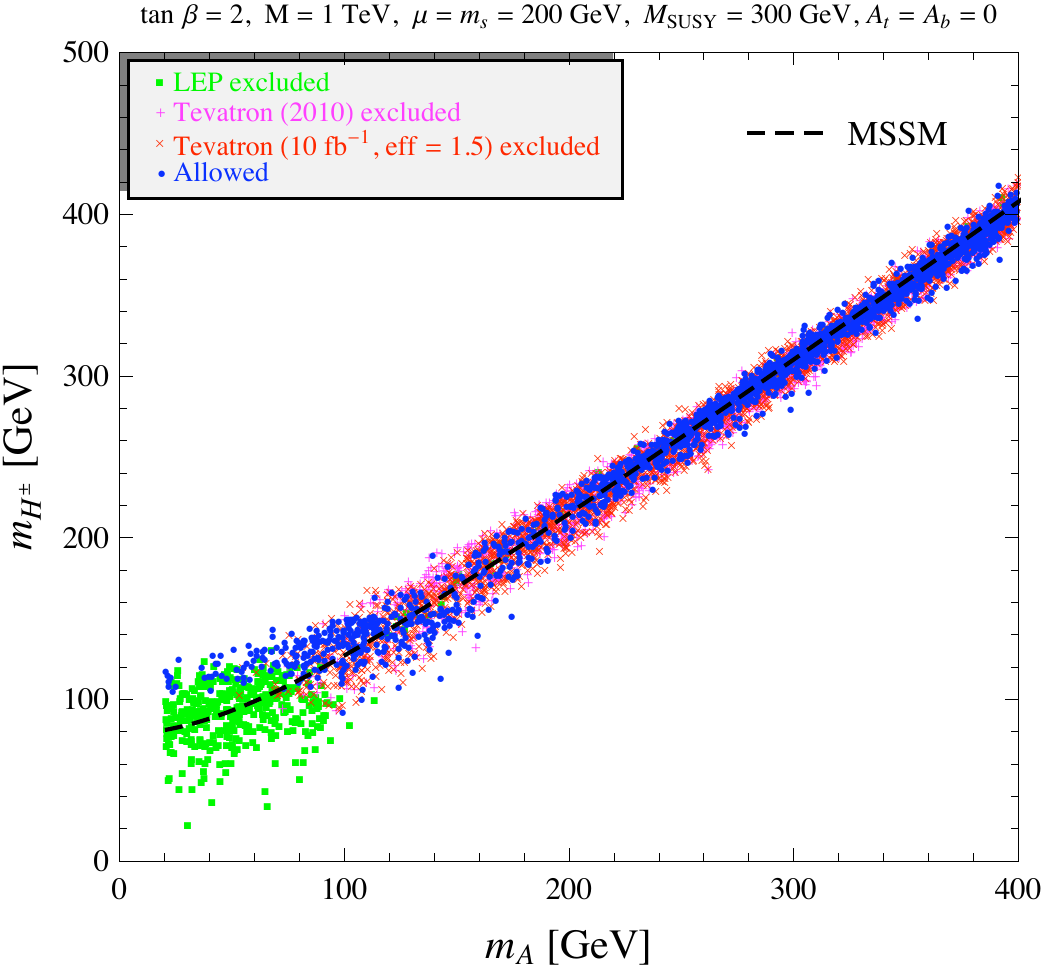}
\hspace{3mm}
\includegraphics[width =7.9cm]{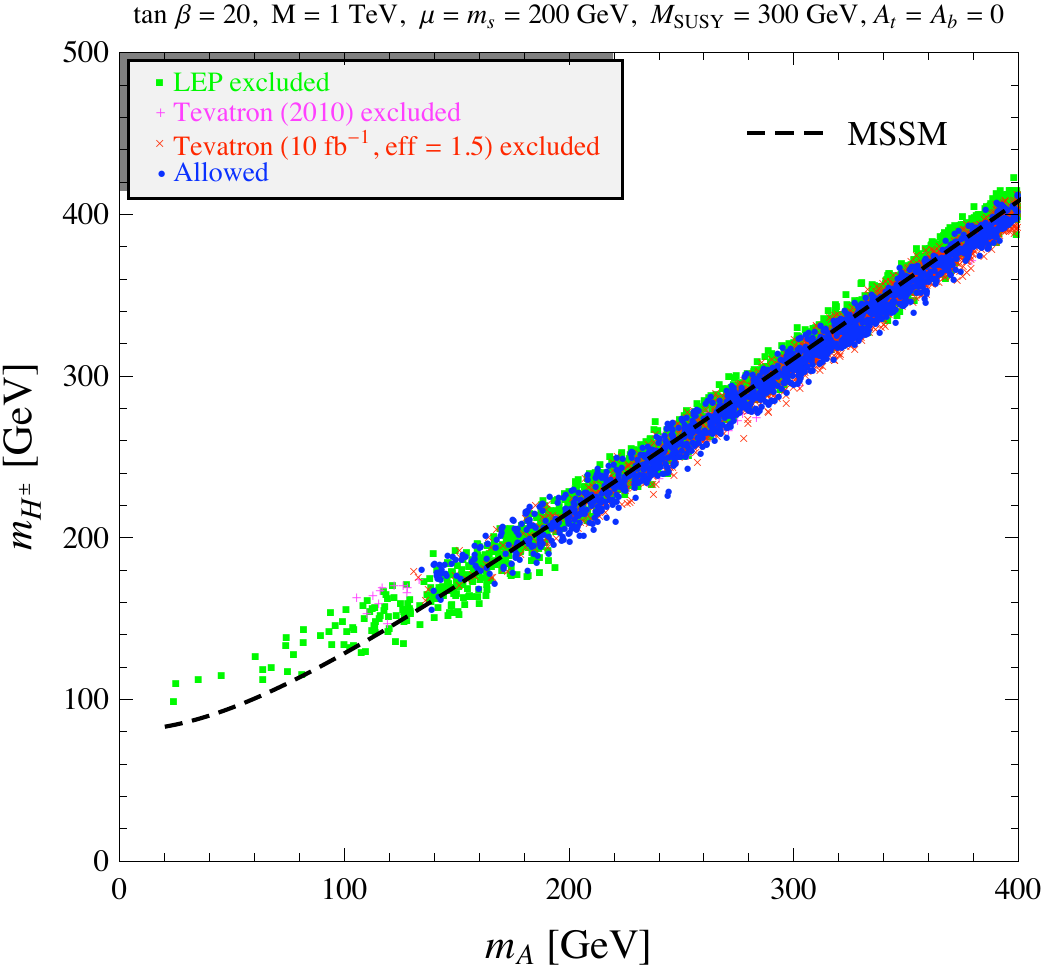}
\end{center}
\caption{\label{fig:mhiggsvsma}{\em $H$ (upper row) and $H^{\pm}$
(lower row) masses as a function of $m_A$, for $\tan \beta=2$ (left
panels) and $\tan \beta=20$ (right panels).  We show the points
excluded by LEP (green), excluded by current Tevatron data (magenta)
and the region that will be proved by the Tevatron in the near future
(red).  The blue points are allowed by all the current experimental
constraints.  The dashed line is the MSSM result for the given SUSY
spectrum.  }}
\end{figure*}
Finally, we show in Fig.~\ref{fig:mhiggsvsma} the masses of the heavy
CP-even and charged Higgs bosons as a function of $m_A$.  The deviations
from the MSSM value are much less dramatic than for $h$.  This is
particularly true in the large $m_{A}$ limit and for large $\tan
\beta$.  Nonetheless, in this region the contribution from the new
physics effects to the masses is of ${\cal O}(10~{\rm GeV)}$, which
cannot be neglected.  For low values of $\tan \beta$ (left plots) we
see that in the region of blue points with low values of $m_A$, both
$m_{H^{\pm}}$ and $m_H$ are above the MSSM value.  Notice that this
effect is more important for $m_H$ than for $m_{H^{\pm}}$.  As a
direct consequence, in the low $\tan \beta$ regime, new \emph{exotic}
channels like $H \to AA$ and $H^{\pm} \to A W^{\pm}$ can be open, with
large BRs, as we will see in the next section.  This does not happen
for $\tan \beta=20$ since, as stated before, there are no allowed
(blue and red) points with $m_A$ below $140~{\rm GeV}$ and the mass
splittings do not allow the previous decay modes.

Having analyzed the modifications in the spectra due to the
higher-dimension operators, we will devote the next section to study
the collider phenomenology of these models.

\section{BMSSM Collider Phenomenology}
\label{sec:results}

In this section we study the phenomenology of the BMSSM Higgs sector,
including all of the effects and constraints described in Section
\ref{sec:review}.  We consider the low and large $\tan \beta$ cases
separately.

\subsection{Low $\tan\beta$ searches: general features}
\label{sec:lowtb}

We start with the low $\tan \beta$ regime, fixing $\tan \beta=2$.
As we have described in the previous section, the main modification
introduced by the higher-dimension operators is to shift the Higgs
spectrum with respect to the MSSM one.  However, the couplings of the
Higgs bosons also get corrected.  The combination of these two
effects can give rise to sizable modifications in both the Higgs
production cross sections and the branching fractions.
 
We compute the production cross sections in the following way.  For
the Higgs-strahlung and vector boson fusion processes, we simply scale
the corresponding SM cross section by the (square of) the
Higgs-$W$-$W$ coupling in our scenario, normalized to the SM coupling,
i.e. by the effective coupling as defined in Eq.~(\ref{geff}) (for all
practical purposes this ratio coincides with the normalized
Higgs-$Z$-$Z$ coupling~\footnote{The difference between $g^2_{h/H WW}$
and $g^2_{h/HZZ}$ arises only from the custodially-violating
higher-dimension operators, and was shown in \cite{Carena:2009gx} to
be numerically negligible.}).  For the gluon fusion cross section, we
shall argue that the NLO K-factor in our scenario is expected to agree
with the NLO K-factor in the SM within 20\%.  This implies that to
this accuracy
\bea
\label{eq:ggheff}
\frac{\sigma^{\rm NLO}(gg \rightarrow h)}{\sigma^{\rm NLO}_{{\rm
SM}}(gg \rightarrow h)} &\approx& 
\frac{\Gamma^{\rm LO}(h \rightarrow gg)}{\Gamma^{\rm LO}_{{\rm
SM}}(h \rightarrow gg)}~,
\eea
since the ratio of cross sections equals the ratio of widths at
leading order in
$\alpha_{s}$~\cite{Georgi:1977gs,Spira:1995rr,Dawson:1996xz}.  The
right-hand side of Eq.~(\ref{eq:ggheff}) is computed using our
modified version of HDECAY~\cite{Djouadi:1997yw}, which includes the
tree-level expressions for masses and couplings in the presence of the
higher-dimension operators.

The K-factor in our scenario differs from the SM one in two respects.
First, the contribution to the gluon fusion cross section from bottom
loops cannot be neglected, specially in the large $\tan \beta$ regime.
Second, one has to consider the presence of a relatively light SUSY
spectrum.  We discuss separately these two effects.  To assess the
impact of the bottom loop we use the code HIGLU~\cite{Spira:1995mt},
that includes both the LO and NLO results for both the SM and the
MSSM~\cite{Spira:1995rr} (but, at present, does not include SUSY
particles in the loop), to compute the K-factors in these two models.
We find that at low $\tan\beta$ and for a wide range of Higgs masses,
the NLO K-factors for $h$, $H$ and $A$ coincide within 5\% with the SM
NLO K-factor for a Higgs of the corresponding mass.  At larger
$\tan\beta$ ($\sim 30$) the differences are larger, as expected, but
still smaller than about 20\%.  We expect that the same will hold in
our extended SUSY scenarios.  The changes in the NLO K-factor due to
relatively light sparticles in the loop, again in the MSSM context,
were studied in \cite{Spira:1997dg}, where the effect was found to be
less than $3 \%$ for $\tan \beta=1.5$.  Therefore, we conclude that at
low $\tan\beta$ Eq.~(\ref{eq:ggheff}) holds to an accuracy of better
than $10 \%$, and allows us to obtain a sufficiently precise estimate
for the NLO gluon fusion cross section in our scenario.  Note that
this uncertainty is below the one obtained by comparing the NLO and
NNLO/NNLL results in the SM calculation
\cite{Harlander:2002wh,Anastasiou:2002yz,Ravindran:2003um,Catani:2003zt,Anastasiou:2008tj,deFlorian:2009hc}.
It is also important to note that the bulk of the effects of the light
SUSY spectrum is taken into account in the LO cross section, and that
these effects are fully implemented in HDECAY, which is used to
compute the right-hand side of Eq.~(\ref{eq:ggheff}).  This also
includes radiative effects that correct the bottom Yukawa coupling,
which can be important at large $\tan\beta$~\cite{Carena:1999py}.
\begin{figure*}[!htp]
\begin{center}
\includegraphics[width=7.9cm]{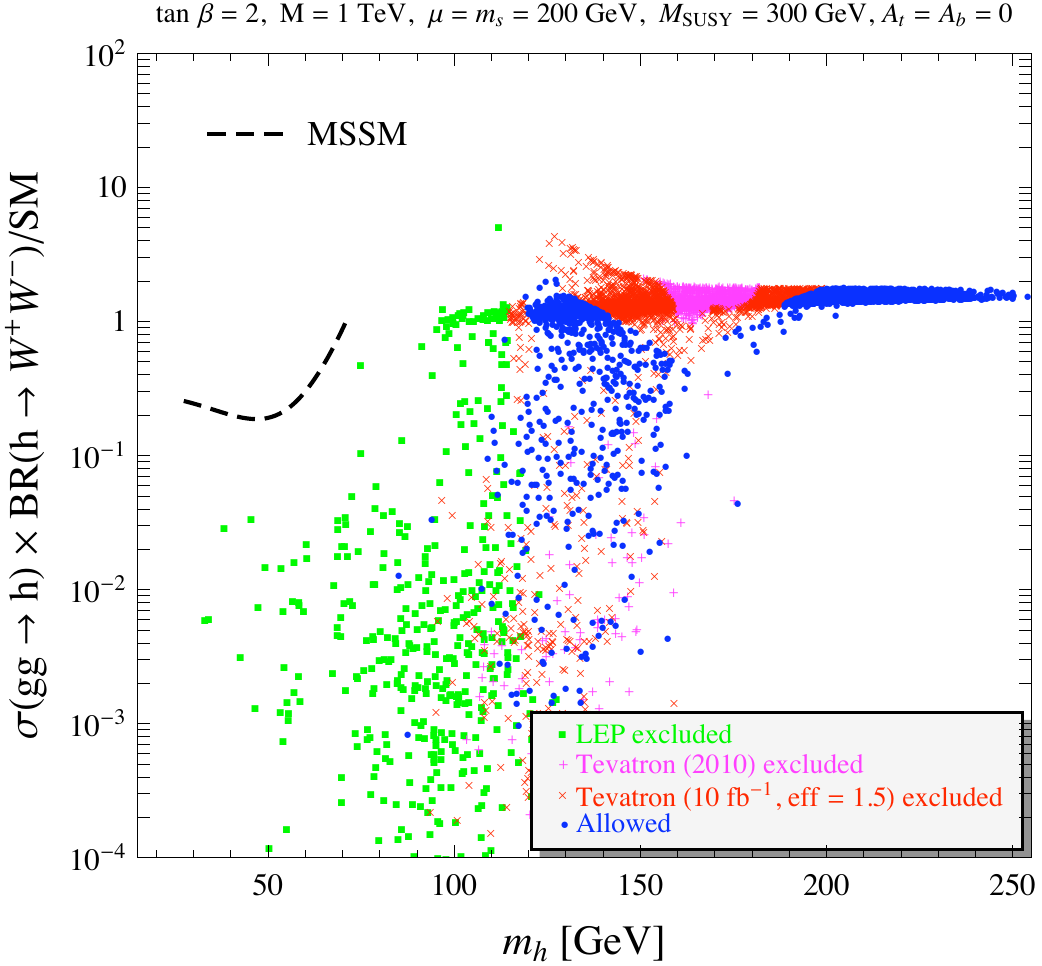}
\hspace{3mm}
\includegraphics[width=7.9cm]{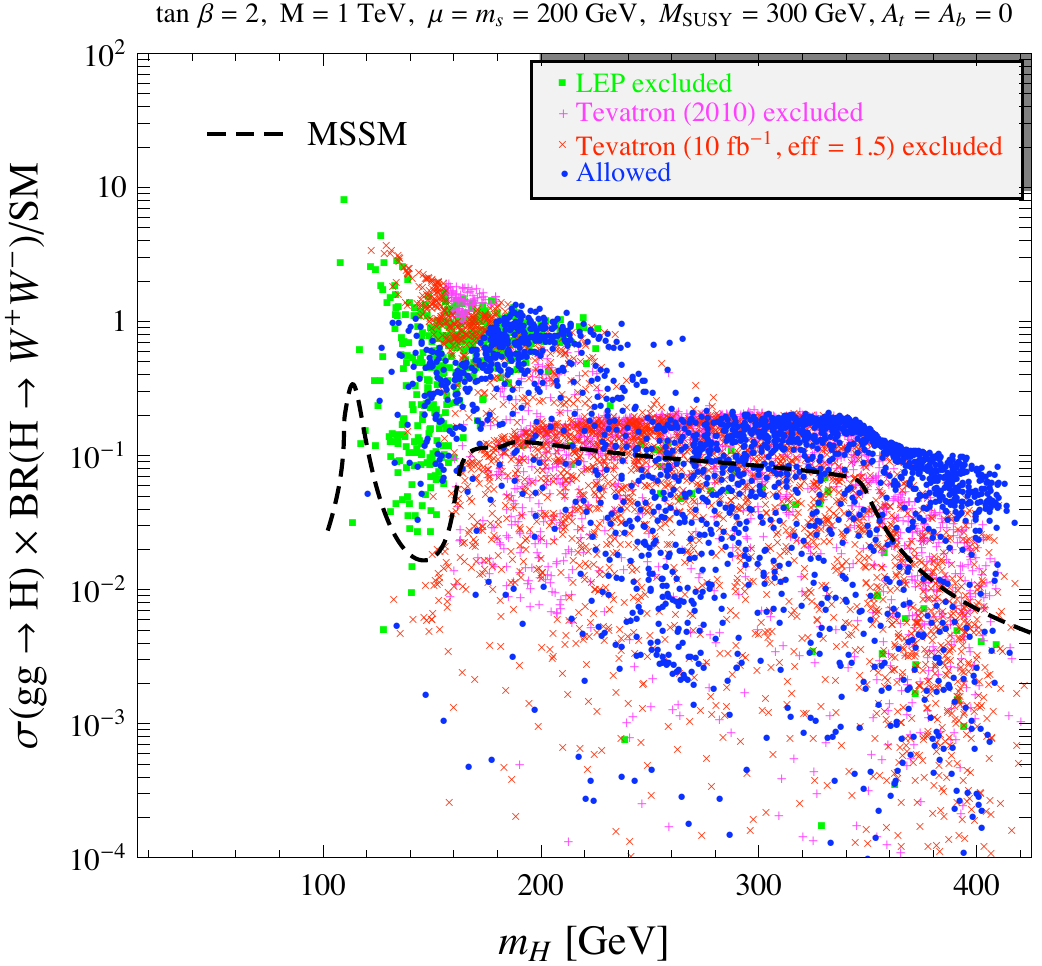}
\end{center}
\vspace{-0.5cm} 
\caption{\label{fig:ggtohtoWWtb2}{\em Production cross
section by gluon fusion times branching ratio into W boson pairs, for
$h$ (left panel) and $H$ (right panel), normalized to the SM result,
as a function of the corresponding Higgs mass, for $\tan \beta=2$.  We
show the points excluded by LEP (green), excluded by current Tevatron
data (magenta) and the region that will be probed by the Tevatron in
the near future (red).  The blue points are allowed by all the current
experimental constraints.  The dashed line corresponds to the MSSM
prediction for the given SUSY spectrum.  }}
\end{figure*}
We discuss next a number of general features regarding the BMSSM Higgs
signals.  In both the SM and the MSSM, the dominant decay channel for
a Higgs boson whose mass is greater than 140 GeV is into W pairs.
Therefore, an important observable at a hadron collider is the
production cross section times the branching fraction in the WW
channel.  In Fig.~\ref{fig:ggtohtoWWtb2} we show this quantity for $h$
and $H$, normalized to the SM result, as a function of the
corresponding Higgs mass.  In the left panel we clearly see the
Tevatron exclusion in the $h \rightarrow WW$ channel: the V-shaped
magenta and red regions around $m_{h} \sim 160~{\rm GeV}$ correspond
to the \emph{stripes} that were already discussed in
Section~\ref{sec:spectra} (see Figs.~\ref{fig:mlvsma} and
\ref{fig:mhvsml}).  We stress that the blue points with $m_h$ above
$180~{\rm GeV}$ have a slightly enhanced $WW$ signal compared to the
SM. In turn, this mass range will be explored at the LHC via the $h
\to ZZ \to 4l$ channel (recall that, for all practical purposes, the
CP-even Higgs normalized couplings to $WW$ and $ZZ$ are the same,
hence the plot can be directly applied to the $ZZ$ channel).  Thus, an
enhanced signal in this region is an interesting feature: for these
points, the Higgs cannot escape detection.  For the blue points with
$WW$ signal reduced by a factor of 10 or more ($m_h < 160~{\rm GeV}$),
one may have to rely on other search channels.

Note that this figure exhibits currently allowed (blue and red) points
with $m_{h}$ below the LEP bound of $114.4~{\rm GeV}$.  These
correspond to models where the coupling to gauge bosons is below the
SM value.  We also notice a group of red points whose signal is around
the SM value, and with a mass slightly above the LEP bound
($114.4~{\rm GeV} \le m_h \lesssim 120~{\rm GeV}$): these are within
the Tevatron reach in the $h \to b\bar{b}$ channel, assuming an
accumulated luminosity of $10~{\rm fb}^{-1}$ per experiment and a 50\%
efficiency improvement in this channel.

Turning our attention to the right panel of
Fig.~\ref{fig:ggtohtoWWtb2}, we see that in the case of $H$ it is hard
to differentiate regions where a single color is predominant, as was
possible in the left panel.  We can identify a mostly green region
with $m_{H}$ above the LEP bound (and above the MSSM curve).  These
points are excluded by the LEP bound on $m_{h}$ rather than on
$m_{H}$, and serve as a reminder that the constraints may come from
observables not related to those shown in a given plot.  This is not
to say that there are no points where the exclusion is through $H$
directly instead of $h$: for instance, the magenta and red points in
the upper left side of the plot correspond to the V-shape exclusion
from $H \to WW$ at the Tevatron.  This is the only region where the
signal is enhanced with respect to both the SM and the MSSM. These
correspond to models where $H$ is SM-like, while $h$ decays mainly
into $b \bar{b}$ and $\tau \bar{\tau}$.

Aside from the $WW$ channel, there are other important decay modes for
light Higgs bosons, in particular $b \bar{b}$, $\tau \bar{\tau}$ and
$\gamma \gamma$.  In the first case, the huge QCD backgrounds render
this channel very difficult to measure at a hadron collider.  This
does not mean, however, that this decay mode is completely useless.
For instance, in the Higgstrahlung process, $q \bar{q} \to Z^{*}/W^{*}
\to Z/W + \textrm{Higgs}$, the gauge boson can be fully reconstructed
from its decay modes, and then $\textrm{Higgs} \to b \bar{b}$ becomes
a feasible option.  Another example to search for a SM-like Higgs
boson decaying into $b\bar{b}$ is the Higgs associated production
together with a top quark pair.  This has the problem of being quite
challenging at hadron colliders.  The di-photon channel, on the other
hand, constitutes the most promising decay channel for a relatively
light SM-like Higgs at the LHC since, in spite of its tiny BR of
${\cal O} (10^{-3})$, an excellent energy resolution can be achieved
and the background is under good experimental control.  The other
important search channel at the LHC in the low Higgs mass range is the
vector boson fusion with the subsequent decay of the Higgs into a
$\tau\bar{\tau}$ pair.

\begin{figure*}[!hbt]
\begin{center}
\includegraphics[width=7.9cm]{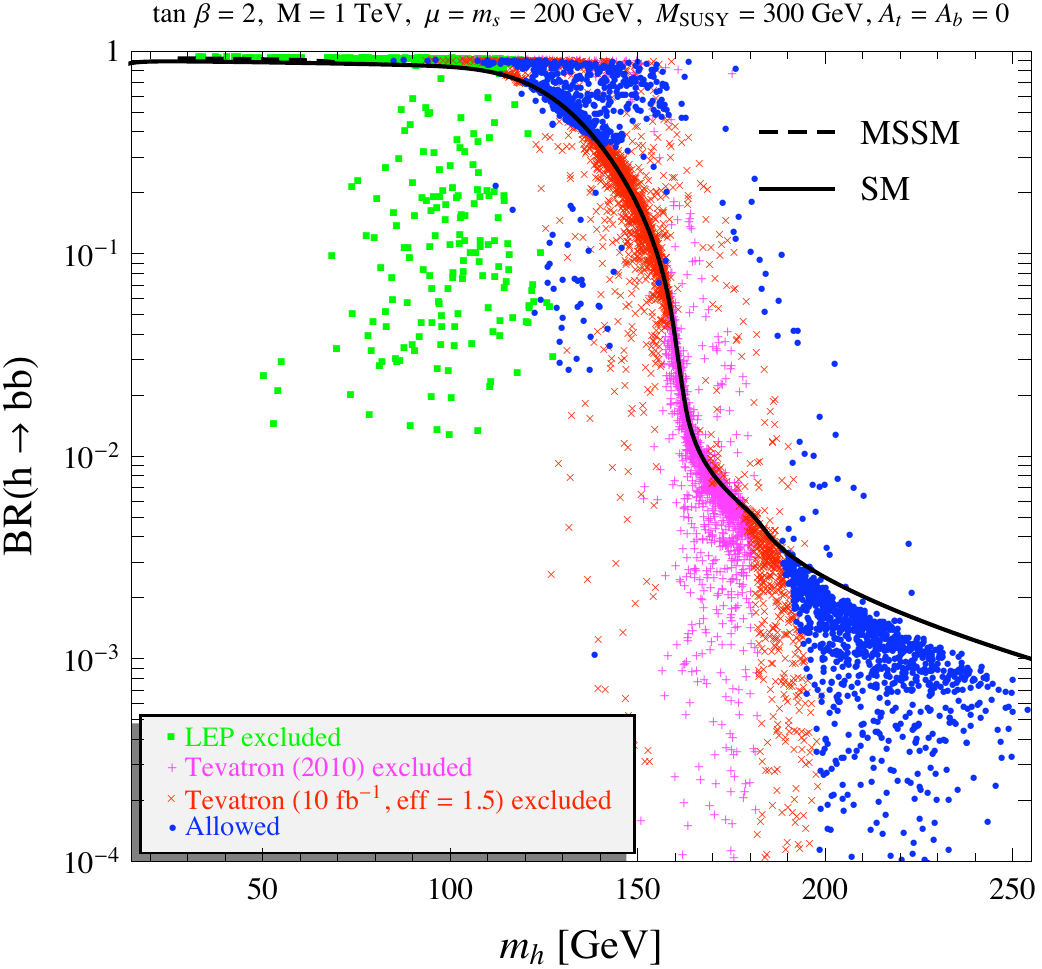}
\hspace{3mm}
\includegraphics[width=7.9cm]{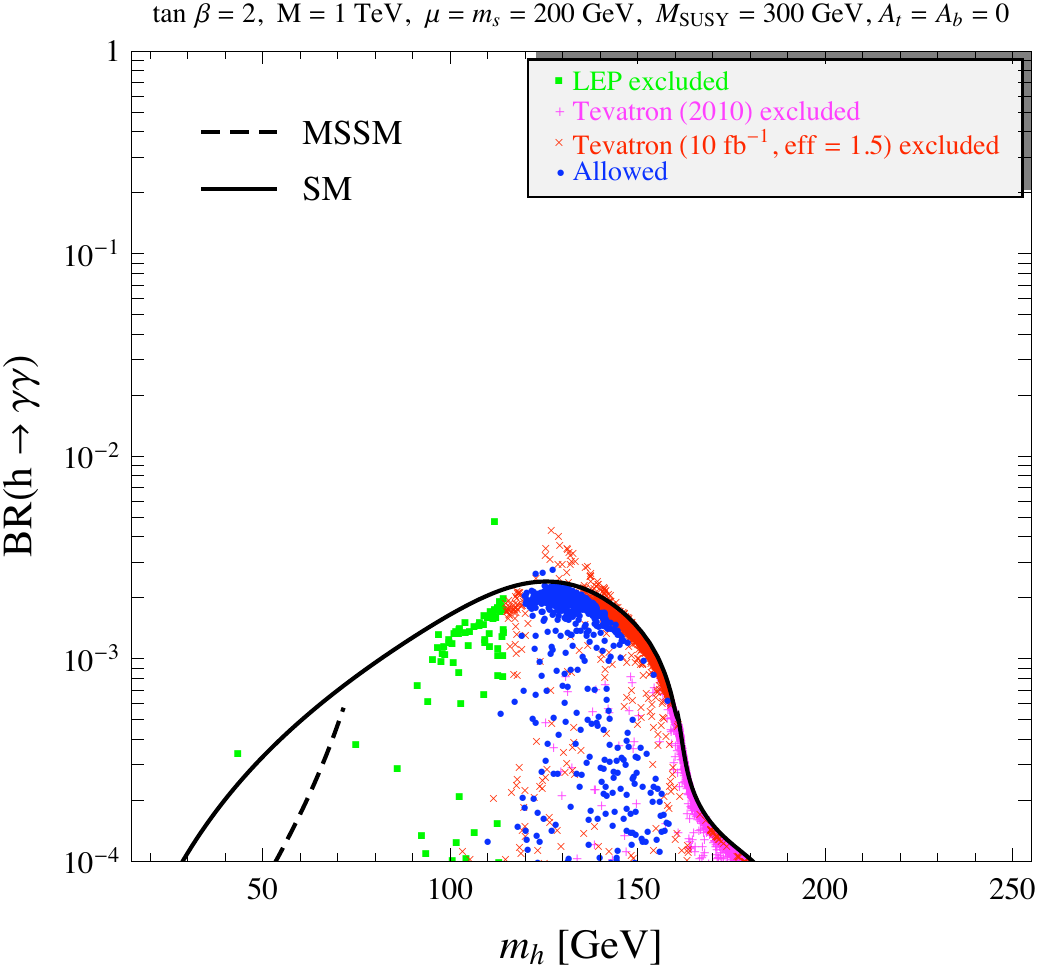}
\end{center}
\vspace{-0.5cm} 
\caption{\label{fig:htobgamtb2}{\em Branching
fractions for $h \to b \bar{b}$ (left panel) and $h \to \gamma \gamma$
(right panel) for $\tan \beta = 2$.  We show the points excluded by
LEP (green), excluded by current Tevatron data (magenta) and the
region that will be probed by the Tevatron in the near future (red).
The blue points are allowed by all the current experimental
constraints.  The solid (dashed) line corresponds to the SM (MSSM)
result.}}
\end{figure*}

In Fig.~\ref{fig:htobgamtb2} we show the branching fraction of $h$
into $b \bar{b}$ (left plot) and $\gamma \gamma$ (right plot), for
$\tan \beta =2$.  Notice that the $b \bar{b}$ channel can be suppressed
with respect to the SM one, as in the blue points with masses in the
$120-150~{\rm GeV}$ range and ${\rm BR}(h \to b\bar{b}) < 10^{-1}$.
This is an interesting feature, since it can lead to enhancements in
other search channels.  One can also see currently allowed (blue and
red) points with BRs into $b\bar{b}$ above the SM curve: those have a
reduced BR into W's, as we have previously identified in
Fig.~\ref{fig:ggtohtoWWtb2}.  In the case of $H$ (not shown here), the
BR into $b\bar{b}$ is typically higher than the SM value.  With
respect to the MSSM, we find that there is no definite tendency: ${\rm
BR}(H \to b\bar{b})$ can be either increased or suppressed by an order
of magnitude.  It is worth mentioning that the branching fraction in
the $\tau \bar{\tau}$ channel follows closely the $b \bar{b}$
behavior.  This is as expected, since the extended Higgs sectors under
consideration do not distinguish between the down-type fermions, in the
sense that the Yukawa coupling normalized to the SM value is the same
for bottoms and taus, while differences due to the SUSY QCD and top Yukawa
interactions, that arise at loop level, are not significant at small
$\tan\beta$.

Turning our attention to the right panel of Fig.~\ref{fig:htobgamtb2},
we see that most models present a suppressed branching fraction in the
di-photon channel.  However, it is worth noticing the group of points
above the SM curve, where an enhancement of up to a factor of 2 can
be achieved.


%
\begin{figure*}[!t]
\begin{center}
\includegraphics[width=7.9cm]{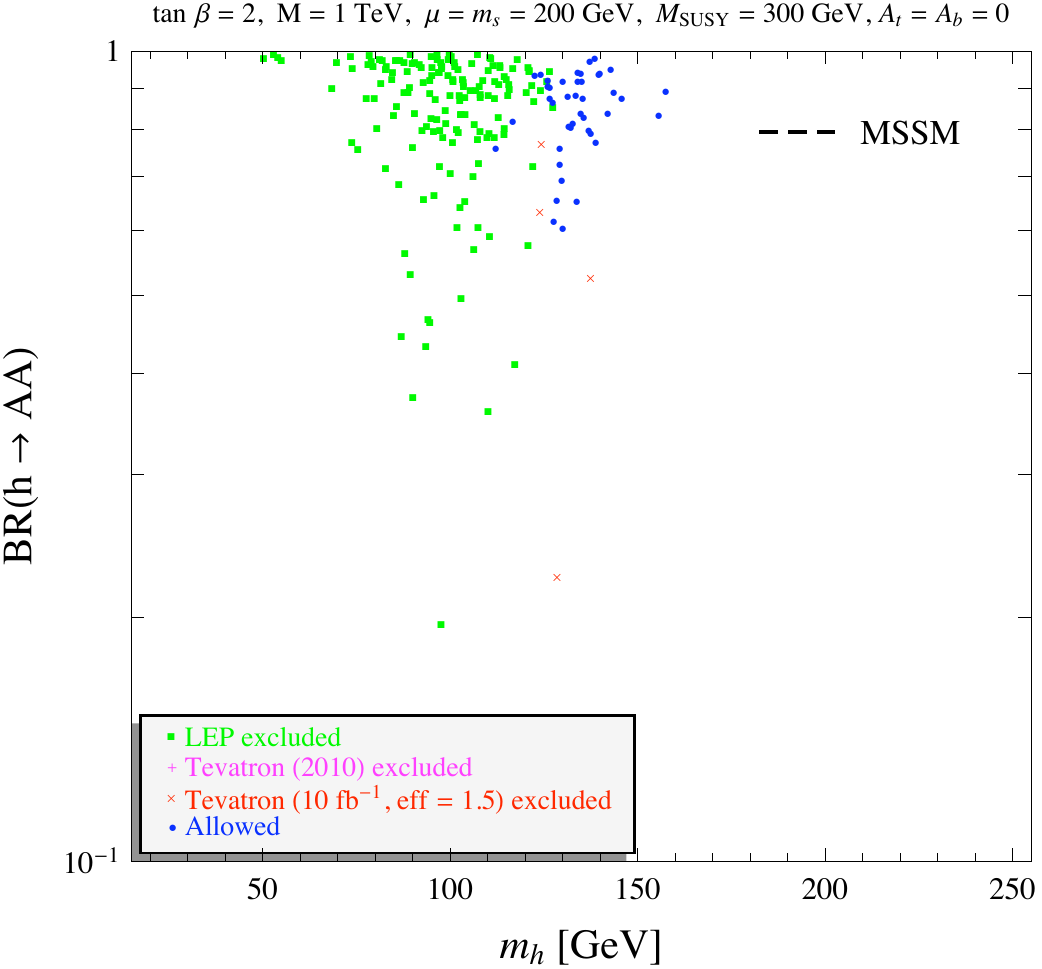}
\hspace{3mm}
\includegraphics[width=7.9cm]{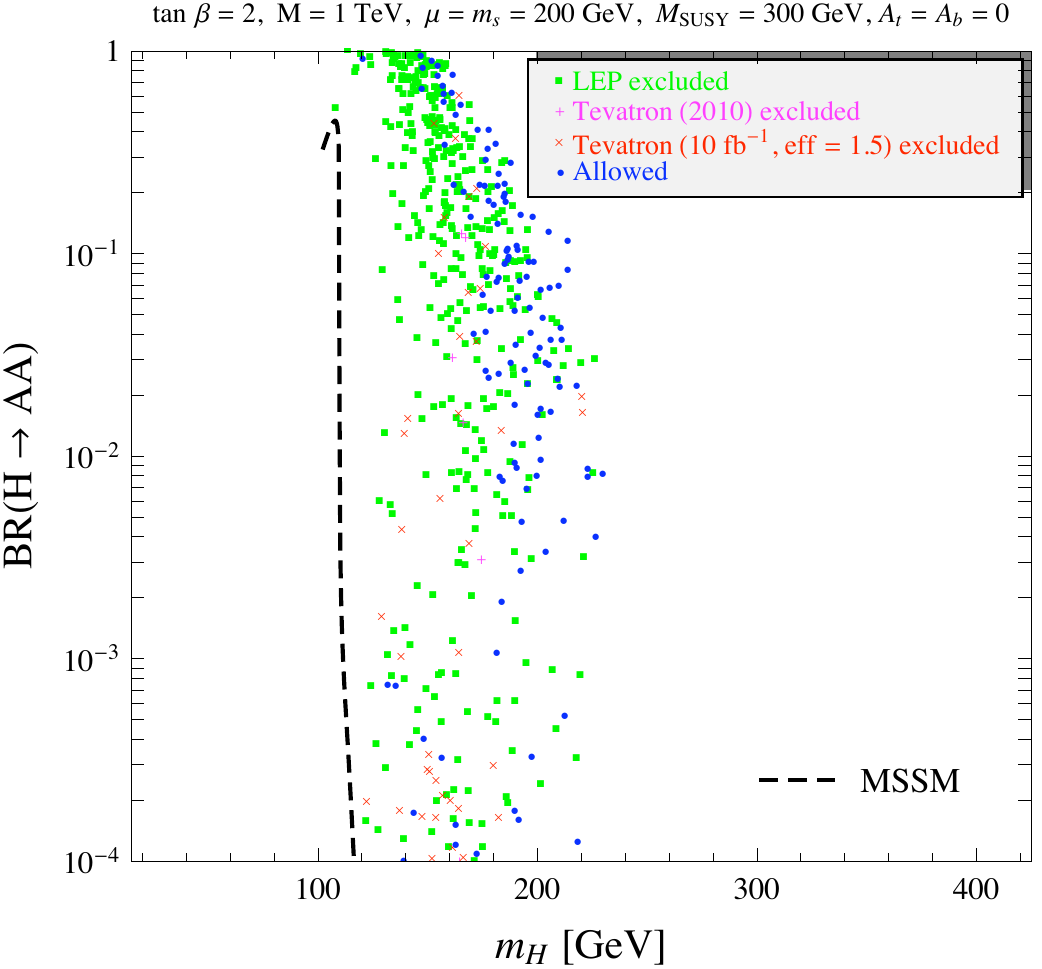}
\end{center}
\caption{\label{fig:htoAA}{\em Branching fractions for $h \to AA$
(left plot) and $H \to AA$ (right plot), for $\tan\beta = 2$.  We show
the points excluded by LEP (green), excluded by current Tevatron data
(magenta) and the region that will be proved by the Tevatron in the
near future (red).  The blue points are allowed by all the current
experimental constraints.  The dashed line corresponds to the MSSM
result for the given SUSY spectrum.}}
\end{figure*}
Decays of the CP-even Higgs bosons into pairs of $A$ bosons can become the
dominant decay mode.  Such a scenario has been previously considered
in the literature (see, for instance, \cite{Carena:2007jk} for a model-independent 
analysis, and \cite{Dermisek:2008uu} for NMSSM studies).
In Fig.~\ref{fig:htoAA} we show the branching fraction of $h$ and $H$
into $AA$, for $\tan \beta =2$.  The left panel shows that the
branching fraction in this channel can reach ${\cal O} (1)$ values,
thus becoming the most relevant decay mode of $h$.  The Tevatron
allowed (blue) points in this figure present a reduced branching
fraction in both the $b \bar{b}$ and the $WW$ channel, and were
already mentioned in the context of Fig.~\ref{fig:htobgamtb2}.  In the
case of $H$ (right panel), the branching fractions vary considerably,
but the $AA$ channel may still become the primary decay mode in some
models.

\begin{figure*}[!t]
\begin{center}
\includegraphics[width=7.9cm]{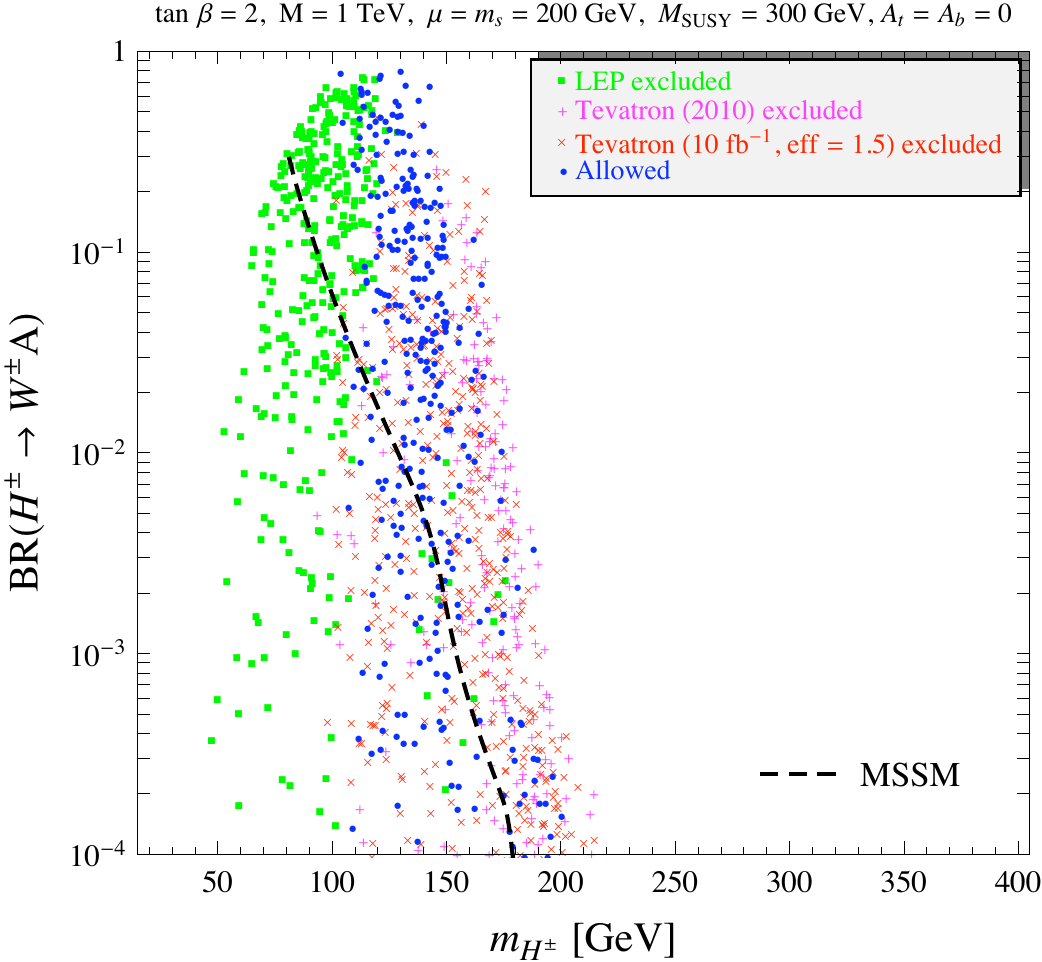}
\hspace{3mm}
\includegraphics[width=7.9cm]{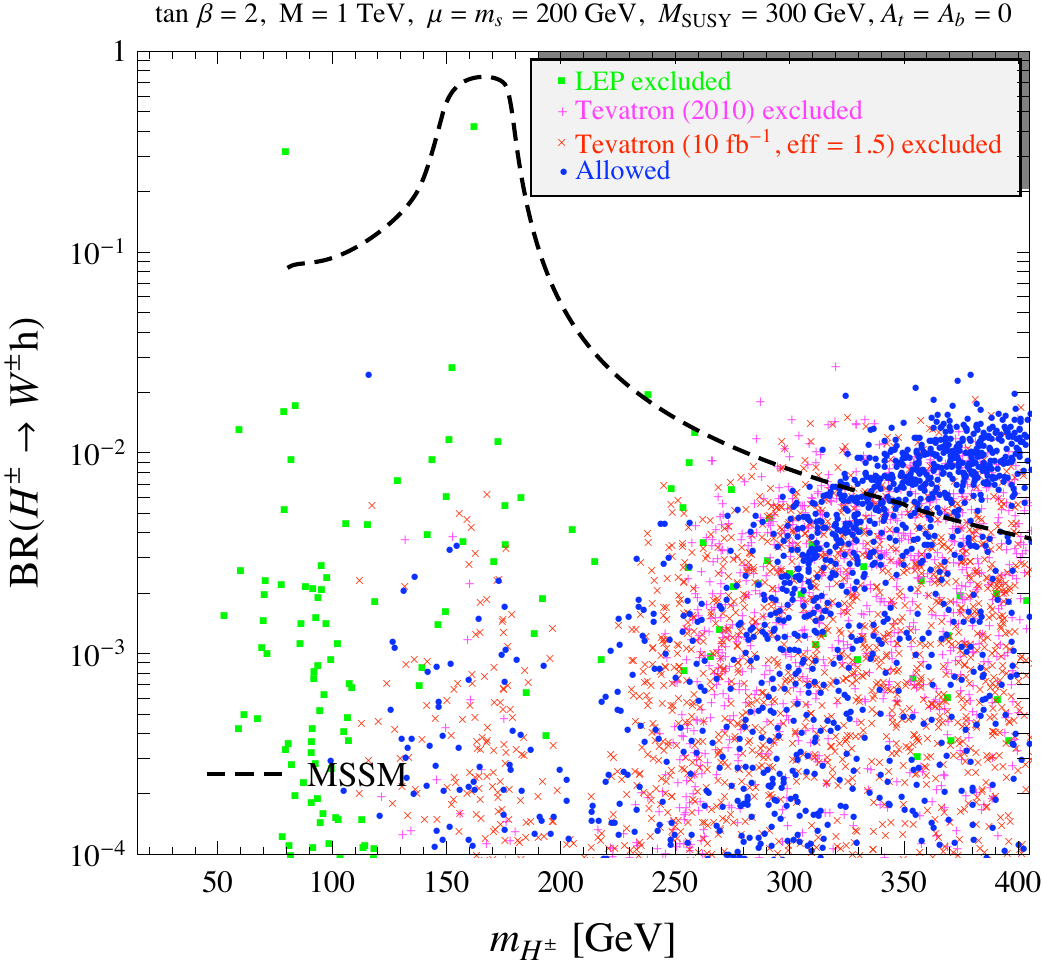}
\end{center}
\caption{\label{fig:Ctb2}{\em $H^{\pm} \to A W^{\pm}$ (left panel) and
$H^{\pm} \to h W^{\pm} $ (right panel), for $\tan\beta = 2$.  The
dashed line corresponds to the MSSM result for the given SUSY
spectrum.}
}
\end{figure*}
The main modification to the decay phenomenology of $A$ and $H^{\pm}$
with respect to the MSSM is due to the shift in the overall Higgs
spectrum.  The channels that change the most are those that involve a
Higgs decaying into either a pair of Higgs bosons, or a Higgs boson
plus a gauge boson.  As an example of the latter, we take the decay of
$H^{\pm}$ into a W boson and a neutral Higgs.  In the MSSM, since $h$
tends to be rather light, one has that $H^{\pm} \to h W^{\pm}$ can be
an important decay channel.  On the other hand, since $m_A$ and
$m_{H^\pm}$ tend to be rather degenerate, one finds that $H^{\pm} \to
A W^{\pm}$ is generally highly suppressed.  On the contrary, in the
context of the BMSSM, one can find points where the mass hierarchy
suffers an inversion, i.e. $m_A$ can be well below $m_h$ and split
from $m_{H^{\pm}}$.  In this case, one finds that $A$ and $h$
interchange their roles with respect to the above described situation
in the MSSM, as can be seen in Fig.~\ref{fig:Ctb2}.  The left panel
shows points where the $A W^{\pm}$ channel has a BR greater than
$0.1$, while the right panel shows that the $h W^{\pm}$ decay mode is
highly suppressed.  In this case, the process $H^{\pm} \to A W^{\pm}
\to b \bar{b} W^{\pm}$ can give rise to an interesting signal at the
LHC, possibly allowing the discovery of two nonstandard Higgs bosons.
Note that we have already encountered another example of an inversion
between $A$ and $h$ in the context of Fig.~\ref{fig:htoAA}: the
potentially open MSSM channels $A \to hh$ and $H \to hh$ are replaced
by $h \to AA$ and $H \to AA$ in the BMSSM context.

Finally, turning our attention to $A$, we have found that the $A \to h
Z$ decay channel is significantly reduced with respect to the MSSM
value $\sim 0.3$ for values of $m_A$ below $250~{\rm GeV}$, due to the
shift in $m_h$ that disfavors this decay mode.~\footnote{Here the
inversion between $h$ and $A$ also takes place, but the $h \to A Z$
decay mode is typically suppressed by a factor of 10 or more with
respect to the dominant decay mode $h \to AA$, when kinematically
allowed.} This reduction brings an enhancement in both the $b \bar{b}$
and $\tau \bar{\tau}$ channels.  As in the MSSM, the former is the
dominant decay channel below the $t\bar{t}$ threshold, and the latter
stays almost constant at about 10\%.
 
\subsection{Low $\tan\beta$ searches: benchmark points}
\label{sec:lowtb:benchmarks}

Up to this point we have analyzed each observable almost independently
of the others.  We would like to understand, however, how the
different features that we have singled out are correlated with each
other.  We shall consider benchmark scenarios currently allowed by LEP
and Tevatron data and explore two possibilities: a) models that can be
probed at 95 \% C.L at the Tevatron in the near future, from now on
referred to as \emph{Tevatron covered} (red) points, and b) models
that are beyond the expected Tevatron reach and will be explored at
the LHC, from now on referred to as \emph{Tevatron uncovered} (blue)
points.  We will also indicate the importance of the $1/M^2$ effects,
as measured by their impact on $m_{h}$.

\subsubsection{Scenarios within the Tevatron reach}
\label{sec:TevatronPoints}

The Tevatron covered models can be divided into three subsets,
according to which channel can exclude the point: $h \to b \bar{b}$,
$h \to WW$ and $H \to WW$.  It is interesting to ask whether a given
model can be probed by more than one channel at the Tevatron.  We
find, however, that the previous subsets are disjoint.  The
disjointness between the subsets probed by $h \to b \bar{b}$ and $h
\to WW$ can be understood in terms of the relevant mass ranges, since
the $b \bar{b}$ search is most sensitive to the $m_h \lesssim 120~{\rm
GeV}$ range, while the di-boson channel probes the region $165 \pm
20~{\rm GeV}$.  In principle, they do not have to be mutually
exclusive, but one would need an enhancement of 3.4 over the SM in the
$h \to WW$ signal in order to probe a $120~{\rm GeV}$ Higgs in this
channel,\footnote{Notice, however, that we have not combined the $h
\to WW$ and $h \to b \bar{b}$ channels.} which is not achievable
within these models: the increase in the $gg \to h \to WW$ cross
section is always below a factor of $2$.  In the case of $h$ and $H$
decaying into W bosons, even if both of them are in the favorable mass
region ($\sim 150-170~{\rm GeV})$, the MSSM sum rule $g^2_{hWW} +
g^2_{HWW}=1$ is valid within 5 \% accuracy, and it is not possible for
$h$ and $H$ to have large enough couplings to W's for both signals to
simultaneously be within the Tevatron reach.  The subsets probed by
the $h \to b \bar{b}$ and $H \to WW$ searches are disjoint because
both processes require a sizable coupling of the Higgs to $W$'s (for
production and decay, respectively), but this does not happen when
$m_{h}$ and $m_{H}$ are sufficiently different, as would be required
for simultaneous searches in these two channels.

\medskip
\noindent
\textit{Point A: MSSM-like scenarios}

\begingroup
\squeezetable
\begin{table}

\begin{center}
\textbf{ POINT A}  
\end{center}
\begin{center}
\begin{tabular}{|c|c|c|c|}
\hline
$m_A ~({\rm GeV})$ & $m_h ~({\rm GeV}) $ & $m_H ~({\rm GeV}) $ & $m_{H^{\pm}} ~({\rm GeV}) $ \\
\hline
239 & 118 & 246 & 245 \\
\hline
$g_{hWW}^2$ & $g_{HWW}^2$  &  $g_{hgg}^2$ & $g_{Hgg}^2$ \\
\hline
0.992   & 0.008 & 1.06  & 0.55 \\
\hline
channel & BMSSM (SM) & channel & BMSSM (SM)  \\ 
\hline
$h \to b \bar{b}$            & 0.78 ~(0.73)  &$ h \to WW $	             & 0.08 ~(0.11) \\
 $ h \to \tau \bar{\tau} $ & 0.08 ~(0.08) & $ h \to \gamma \gamma / 10^{-3}$  & 1.42 ~(2.30) \\

$H \to b \bar{b}$            & 0.15   &
$ H \to WW $	             & 0.22   \\
$ H \to ZZ    $	 	    & 0.11   &
$ H \to hh    $     	    & 0.50   \\
$A \to b \bar{b}$            & 0.89  & $ H^{+} \to t \bar{b}$     & 0.99  \\
$ A \to \tau \bar{\tau} $  & 0.08  & $ A \to Z h$ & 0.24  \\
\hline
\end{tabular}
\end{center}
\caption{ {\em Masses and branching fractions in the BMSSM (and in the
SM for $h$) for point A. We only show the main decay modes.  The
effective couplings $g^{2}_{\phi X}$ were defined in
Eq.~(\ref{geff}). }}
\label{tab:pointA}
\end{table}
\endgroup
We start our analysis with the points that can be probed via the $h
\to b \bar{b}$ decay mode.  We find that these do not differ greatly
from the decoupling limit of the MSSM with rather heavy sparticles
($\sim$ a few TeV).  In this case, the observation of a light SUSY
spectrum (in the few hundred$~{\rm GeV}$ range) would be the smoking
gun of BMSSM physics, since such a light SUSY spectrum would be in
conflict with the LEP limits on the MSSM lightest CP-even Higgs boson.
We illustrate the main features of this subset by showing point
A~\footnote{Following the notation of \cite{Carena:2009gx}, the
coefficients of the effective operators for this point are:
\mbox{$\omega_1=0.23$}, \mbox{$\alpha_1=0.76$}, \mbox{$c_1=-0.41$},
\mbox{$\gamma_1=0.73$}, \mbox{$\beta_1=-0.83$}, \mbox{$c_2=-0.22$},
\mbox{$\gamma_2=0.95$}, \mbox{$\beta_2=0.86$}, \mbox{$c_3=-0.44$},
\mbox{$\gamma_3=-0.84$}, \mbox{$\beta_3=0.95$}, \mbox{$c_4=-0.53$},
\mbox{$\gamma_4=0.44$}, \mbox{$\beta_4=-0.85$}, \mbox{$c_6=-0.31$},
\mbox{$\delta_6=0.93$}, \mbox{$\gamma_6=0.54$}, \mbox{$\beta_6=0.48$},
\mbox{$c_7=0.50$}, \mbox{$\delta_7=0.50$}, \mbox{$\gamma_7=0.78$},
\mbox{$\beta_7=0.98$} .  } in Table~\ref{tab:pointA}, where we include
the mass spectrum and the branching fractions of the most important
decay channels for each Higgs boson.  For reference, in the case of
$h$ we also indicate between parentheses the SM values.  We also note
that for this point, the $1/M^2$ operators contribute about $25~{\rm
GeV}$ to $m_{h}$.

Generically, the branching fractions of $h$ do not deviate much from
the SM ones.  One finds a small increase in the $h \to b \bar{b}$
channel, while ${\rm BR}(h \to \gamma \gamma)$ is slightly suppressed
with respect to the SM (by at most a factor of 3).  Since $g_{hWW}^2 $
and $g_{hgg}^2$ are close to one, the production cross sections by
Higgs-strahlung and gluon fusion are, for all practical purposes, the
same as in the SM. Thus, the change in the signal is given by the
ratio of the branching fractions in our scenario to those in the SM.
For point A, the production rate in $h \to b \bar{b}$ is 6\% above the SM
result.  In this case, the Tevatron could claim a hint on a SM Higgs
boson, while at the LHC the direct detection of $h$ would proceed in
the di-photon channel, since the $gg \to h \to \gamma \gamma$ cross
section is $0.65$ of the SM value.  Some of the remaining Higgs bosons
may also be observed.  For $H$ and $A$, the $H \to h h \to \gamma
\gamma b \bar{b}$ and $A \to Z h \to l l b \bar{b}$ searches provide
the best prospects for discovery \cite{atlasphystdr}.  For a charged
Higgs with a mass above $m_t$, the ATLAS update of 2009
\cite{Aad:2009wy} found that the $tb$ channel is rather challenging
and that the low $\tan\beta$ region cannot be covered.
  
\medskip
\noindent
\textit{Point B: Light Higgs spectra}

\begingroup
\squeezetable
\begin{table}
\begin{center}
\textbf{ POINT B} 
\end{center}
\begin{center}
\begin{tabular}{|c|c|c|c|}
\hline
$m_A ~({\rm GeV})$ & $m_h ~({\rm GeV}) $ & $m_H ~({\rm GeV}) $ & $m_{H^{\pm}} ~({\rm GeV}) $ \\
\hline
101 & 129 & 141 & 135 \\
\hline
$g_{hWW}^2$ & $g_{HWW}^2$  &  $g_{hgg}^2$ & $g_{Hgg}^2$ \\
\hline
0.8 &  0.2 & 1.72 & 0.06 \\
\hline
channel & BMSSM (SM) & channel & BMSSM (SM)  \\ 
\hline
$h \to b \bar{b}  $     & 0.01 ~(0.56) & $h \to \tau \bar{\tau} $     & 0.001 ~(0.06)  \\
$ h \to WW$  & 0.63 ~(0.28) & $ h \to ZZ$  & 0.08 ~(0.04) \\
$ h \to {\rm jets} $          & 0.26 ~(0.06) &$ h \to \gamma \gamma / 10^{-3}$  & 3.97 ~(2.38) \\
$H \to b \bar{b}/ \tau \bar{\tau}$            & 0.84 / 0.09 & $ H \to WW    $	 	    & 0.05    \\
$A \to b \bar{b} / \tau \bar{\tau}$            & 0.89 / 0.09  & $ H^{+} \to \tau \nu_{\tau}$ & 0.87 \\
\hline
\end{tabular}
\end{center}
\caption{ {\em Masses and branching fractions in the BMSSM (and in the
SM for $h$) for point B. 
}}
\label{tab:pointB}
\end{table}
\endgroup
We turn now our attention to the models that can be excluded at the
Tevatron by the $h \to WW$ channel.  Those can be further split into
two categories, according to whether $m_h$ is high ($\gtrsim 170~{\rm
GeV}$) or low ($\lesssim 160~{\rm GeV}$), corresponding to the two red
stripes defined in the context of Fig.~\ref{fig:mlvsma}.  As a general
feature of the lower red stripe, the branching fraction of $h$ into $b
\bar{b}$ can be sizably reduced with respect to the SM, as we pointed
out in the left panel of Fig.~\ref{fig:htobgamtb2}.  This implies that
the remaining channels are enhanced, which is interesting for the $h
\to WW$ and $h \to \gamma \gamma$ decay modes.  We present as an
example point B~\footnote{The Lagrangian parameters for this point
are: \mbox{$\omega_1=0.57$}, \mbox{$\alpha_1=-0.50$},
\mbox{$c_1=-0.24$}, \mbox{$\gamma_1=-0.39$}, \mbox{$\beta_1=0.77$},
\mbox{$c_2=-0.34$}, \mbox{$\gamma_2=-0.53$}, \mbox{$\beta_2=-0.91$},
\mbox{$c_3=-0.80$}, \mbox{$\gamma_3=-0.50$}, \mbox{$\beta_3=-0.49$},
\mbox{$c_4=0.02$}, \mbox{$\gamma_4=-0.34$}, \mbox{$\beta_4=0.59$},
\mbox{$c_6=0.32$}, \mbox{$\delta_6=-0.59$}, \mbox{$\gamma_6=0.34$},
\mbox{$\beta_6=-1$}, \mbox{$c_7=0.57$}, \mbox{$\delta_7=-0.77$},
\mbox{$\gamma_7=0.85$}, \mbox{$\beta_7=-0.89$}.  } in
Table~\ref{tab:pointB}.  Here, one sees that the Higgs spectrum is
relatively light.  $H$ is the heaviest Higgs, while $h$ is lighter
than $H^{\pm}$, but heavier than $A$.  It turns out that for this
point, the $1/M^2$ effects result in a slight net reduction of $m_{h}$
by a couple of GeV.

Since $h$ is SM-like, we give in parentheses the corresponding
branching fractions in the SM. Here we clearly observe that $h$
presents an increase in the gluon fusion cross section, and in the
branching fractions into photons and W bosons, accompanied by a
sizable reduction in the down-type fermion decay modes.  Note that the
$gg \to h \to \gamma \gamma$ and $gg \to h \to WW$ signals are larger
than in the SM by factors of $2.86$ and $3.82$ respectively, which
would facilitate the search of $h$ at the LHC as well.  $H$ decays
mainly into bottoms and taus, and its production cross section by
gluon fusion is strongly reduced with respect to the SM case.  The
most promising discovery channel at the LHC would be $q q H \to q q
\tau \bar{\tau}$, where the signal is reduced with respect to the SM
by a factor of two.  The CP-odd $A$ decays as in the MSSM, while for
the charged Higgs the $\tau \nu_{\tau}$ channel is the dominant one.
 
\medskip
\noindent
\textit{Point C: The heavy CP-even $H$ as the SM-like Higgs}

In the high $m_h$ region that can be probed at the Tevatron in the $h
\to WW$ channel one finds an unusual SUSY spectrum.  Typically, one
runs into the previously mentioned \emph{inversions} between $h$ and
$A$.  Moreover, $h$ can also be heavier than the charged Higgs, which
is a feature that is not present in the region where $m_h$ is below
$150~{\rm GeV}$.  We illustrate this with point C~\footnote{The
Lagrangian parameters for this point are: \mbox{$\omega_1=0.86$},
\mbox{$\alpha_1=-0.70$}, \mbox{$c_1=0.75$}, \mbox{$\gamma_1=0.48$},
\mbox{$\beta_1=0.39$}, \mbox{$c_2=-0.42$}, \mbox{$\gamma_2=-0.64$},
\mbox{$\beta_2=0.86$}, \mbox{$c_3=-0.61$}, \mbox{$\gamma_3=0.82$},
\mbox{$\beta_3=-0.78$}, \mbox{$c_4=-0.49$}, \mbox{$\gamma_4=-0.82$},
\mbox{$\beta_4=-1$}, \mbox{$c_6=-0.05$}, \mbox{$\delta_6=-0.41$},
\mbox{$\gamma_6=-0.56$}, \mbox{$\beta_6=0.97$}, \mbox{$c_7= 0.88$},
\mbox{$\delta_7=0.75$}, \mbox{$\gamma_7=0.80$}, \mbox{$\beta_7=0.56$}.
} in Table~\ref{tab:pointC}.  For this point, the $1/M^2$ operators
contribute about $30~{\rm GeV}$ to $m_{h}$.

\begingroup
\squeezetable
\begin{table}
\begin{center}
\textbf{ POINT C} 
\end{center}
\begin{center}
\begin{tabular}{|c|c|c|c|}
\hline
$m_A ~({\rm GeV})$ & $m_h ~({\rm GeV}) $ & $m_H ~({\rm GeV}) $ & $m_{H^{\pm}} ~({\rm GeV}) $ \\
\hline
135 & 174 & 186 & 164 \\
\hline
$g_{hWW}^2$ & $g_{HWW}^2$  &  $g_{hgg}^2$ & $g_{Hgg}^2$ \\
\hline
0.11 & 0.89 &  1.05 & 0.65 \\
\hline
channel & BMSSM (SM) & channel & BMSSM (SM)  \\ 
\hline
$h \to b \bar{b}  $     & 0.12 ~(0.01) & $ h \to WW$  & 0.84 ~(0.96) \\
$ H \to WW    $	 	    & 0.81 ~(0.82)  &
$ H \to ZZ    $     	    & 0.17 ~(0.17)  \\
$A \to b \bar{b} $           & 0.90 \ & $A \to \tau \bar{\tau}$  &0.10 \\
$ H^{+} \to \tau \nu_{\tau} $ & 0.59  &
$ H^{+} \to t \bar{b}$   & 0.38   \\
\hline
\end{tabular}
\end{center}
\caption{ {\em Masses and branching fractions in the BMSSM (and in the
SM for $h$ and $H$) for point C. }}
\label{tab:pointC}
\end{table}
\endgroup

Here we see that the two CP-even Higgs bosons have masses well above
the maximum value for $m_h$ that can be obtained in the $m_h$ max
scenario within the MSSM context.  In this case, it makes sense to
compare both $h$ and $H$ with the SM Higgs.  For this particular
point, $h$ has not been yet excluded by the Tevatron search
since it is not SM-like and its branching fraction into $WW$ is
somewhat suppressed.  $H$ can be discovered by the LHC in the $ZZ \to
4 l$ mode, since the signal normalized to the SM value, is $0.65$.  We
note that here both $m_{h}$ and $m_{H}$ are near the region where the
$WW$ channel opens up leading to a suppression in the sensitivity of
the $ZZ$ search mode at the LHC. We recall that such a heavy SM-like
$H$ is not a feature of the MSSM, being a unique characteristic of the
BMSSM Higgs sector.  The CP-odd $A$ decays almost entirely to bottom
and tau pairs, while the charged Higgs has sizable decays into both
the $\tau \nu_{\tau}$ and $t \bar{b}$ channels.

\medskip
\noindent

The last subset of the Tevatron covered points corresponds to those
than can be probed by the $H \to WW$ search, for which the $gg \to H
\to WW$ signal goes between $0.4 - 4$ times the SM value.  In such
scenarios, $h$ and $A$ decay mostly into bottoms and taus.  In some
cases the $h \to \gamma \gamma$ or $q q H \to q q \tau \tau$ signal
might be observable at the LHC. The charged Higgs is relatively light
(always below $200~{\rm GeV} $) and will decay almost $100 \% $ of the
time into $\tau \nu_{\tau}$ for masses below $160~{\rm GeV}$, and in
$t \bar{b}$ for the remaining points.  The $H^{\pm} \to h W^{\pm}$
decay mode is closed for kinematical reasons, as we already know from
Fig.~\ref{fig:Ctb2}.  In addition, when $m_A$ is light, the $h \to AA$
and $H^{\pm} \to A W^{\pm}$ channels might become important.  Since
this also happens with the Tevatron uncovered points, we will defer
further comments and the study of a suitable benchmark point for the
next subsection.

\subsubsection{LHC searches}
\label{sec:LHCPoints}

Regarding the Tevatron uncovered points, we can also split them
into two disjoint subsets, corresponding to each of the blue stripes
in Figs.~\ref{fig:mlvsma} or \ref{fig:mhvsml}: we will refer to them
as low mass ($m_h \lesssim 140~{\rm GeV}$) and high mass (above
$190~{\rm GeV}$) regions.  In the high $m_h$ case, one can make a
further distinction according to whether $m_A$ is below or above
$160~{\rm GeV}$.  Again, we illustrate the possibilities with a few
benchmark points.

\medskip
\noindent
\textit{Point D: Two peaks in the $ZZ \to 4l$ signal}

\begingroup
\squeezetable
\begin{table}
\begin{center}
\textbf{ POINT D} 
\end{center}
\begin{center}
\begin{tabular}{|c|c|c|c|}
\hline
$m_A ~({\rm GeV})$ & $m_h ~({\rm GeV}) $ & $m_H ~({\rm GeV}) $ & $m_{H^{\pm}} ~({\rm GeV}) $ \\
\hline
184 & 204 & 234 & 203 \\
\hline
$g_{hWW}^2$ & $g_{HWW}^2$  &  $g_{hgg}^2$ & $g_{Hgg}^2$ \\
\hline
0.3 & 0.7 &  1.39 & 0.36 \\
\hline
channel & BMSSM (SM) & channel & BMSSM (SM)  \\ 
\hline
$h \to WW $     & 0.73 ~(0.72) & $ h \to ZZ$  & 0.25 ~(0.27) \\
$ H \to WW    $	 	    & 0.70 ~(0.71)  &
$ H \to ZZ    $     	    & 0.29 ~(0.29)  \\
$A \to b \bar{b}$             & 0.87 &
$ H^{+} \to t \bar{b} $ & 0.99 \\
\hline
\end{tabular}
\end{center}
\caption{ {\em Masses and branching fractions in the BMSSM (and in the
SM for $h$ and $H$), for point D.  }}
\label{tab:pointD}
\end{table}
\endgroup
General features of the high $m_{h}$, high $m_{A}$ case are: increased
$gg \to h$ cross section with respect to the SM, and negligible (below
2\%) changes in the $h \to WW/ZZ$ decay modes.  Regarding $H$, one has
that the signal in the $gg \to H \to WW/ZZ$ channel is always
suppressed with respect to the SM. As an example, we show point
D~\footnote{The Lagrangian parameters for this point are:
\mbox{$\omega_1=1$}, \mbox{$\alpha_1=-1$}, \mbox{$c_1=0.57$},
\mbox{$\gamma_1=-0.77$}, \mbox{$\beta_1=-0.44$}, \mbox{$c_2=-0.38$},
\mbox{$\gamma_2=-0.38$}, \mbox{$\beta_2=0.72$}, \mbox{$c_3=-0.80$},
\mbox{$\gamma_3=0.88$}, \mbox{$\beta_3=-0.52$}, \mbox{$c_4=-0.94$},
\mbox{$\gamma_4=-0.89$}, \mbox{$\beta_4=-0.34$}, \mbox{$c_6=0.30$},
\mbox{$\delta_6=-0.47$}, \mbox{$\gamma_6=0.81$},
\mbox{$\beta_6=-0.41$}, \mbox{$c_7= -0.71$}, \mbox{$\delta_7=-0.70$},
\mbox{$\gamma_7=0.38$}, \mbox{$\beta_7=0.62$}.  } in
Table~\ref{tab:pointD}.  Here, the $1/M^2$ operators contribute about
$40~{\rm GeV}$ to $m_{h}$.  Given the features of this point, it again
makes sense to compare both CP-even Higgs bosons with the SM.

The rise in the mass of $h$ automatically closes the $H \to hh$, $A
\to h Z$ and $H^{\pm} \to h W^{\pm}$ decay modes, which could be
important in the MSSM case.  This picture can suffer some alterations
if $h$ stays around $200~{\rm GeV}$, while the rest of the Higgs
bosons attain values around $400~{\rm GeV}$, since this will open not
only the previously mentioned channels, but possibly also decays into
sparticles.  In such a case, one would run into a sort of \emph{MSSM
decoupling limit}, but with a mass for the lightest Higgs which is
unattainable within the MSSM. Concentrating on point D, we emphasize
that both $h$ and $H$ couple in a sizable way to the electroweak gauge
bosons, and thus the measurement of both couplings will permit a
detailed study of the EWSB mechanism, as it arises from a 2HDM. One
possibility to discover these CP-even Higgs bosons would be to search
for two isolated peaks in the $ZZ \to 4l$ golden mode.  Notice that
the branching fractions of both $h$ and $H$ are very close to their SM
counterparts, while there is a difference in the gluon fusion
production cross section.  Since the $gg \to H \to ZZ$ signal can be
sizably suppressed with respect to the SM, the direct detection of $H$
in this channel might not be feasible.  Nevertheless, a large number
of models similar to point D would in fact present two clear peaks in
the di-lepton invariant mass distribution.  We also note that these
points have a $gg \to h \to WW/ZZ$ signal that can be 20-70 \% larger
than the SM value.

\medskip
\noindent
\textit{Point E: Non-SM-like Higgs with a clear di-boson signal}

\begingroup
\squeezetable
\begin{table}
\begin{center}
\textbf{ POINT E} 
\end{center}
\begin{center}
\begin{tabular}{|c|c|c|c|}
\hline
$m_A ~({\rm GeV})$ & $m_h ~({\rm GeV}) $ & $m_H ~({\rm GeV}) $ & $m_{H^{\pm}} ~({\rm GeV}) $ \\
\hline
134 & 181 & 205 & 165 \\
\hline
$g_{hWW}^2$ & $g_{HWW}^2$  &  $g_{hgg}^2$ & $g_{Hgg}^2$ \\
\hline
0.03 & 0.95 &  0.79 & 0.99 \\
\hline
channel & BMSSM (SM) & channel & BMSSM (SM)  \\ 
\hline
$h \to b \bar{b}$            & 0.23 ~(0.005)  &
$ h \to \tau \bar{\tau} $	             & 0.03 ~(0.0005)  \\
$h \to WW $     & 0.68 ~(0.92) & $ h \to ZZ$  & 0.04 ~(0.07) \\
$ H \to WW    $	 	    & 0.72 ~(0.73)  &
$ H \to ZZ    $     	    & 0.27 ~(0.27)  \\
$A \to b \bar{b}$            & 0.89 &
$A \to \tau \bar{\tau}$   &0.10 \\
 $ H^{+} \to t \bar{b} $    & 0.57  &
$ H^{+} \to \tau \nu_{\tau}$   & $0.40$    \\
\hline
\end{tabular}
\end{center}
\caption{ {\em Masses and branching fractions (and in the SM for $h$
and $H$) for point E. }}
\label{tab:pointE}
\end{table}
\endgroup
One interesting example of a point where $m_h$ is still high, but
$m_A$ is below $160~{\rm GeV}$ is given in point E~\footnote{The
Lagrangian parameters for this point are: \mbox{$\omega_1=1$},
\mbox{$\alpha_1=-1$}, \mbox{$c_1=0.77$}, \mbox{$\gamma_1=-0.89$},
\mbox{$\beta_1=-0.97$}, \mbox{$c_2=-0.44$}, \mbox{$\gamma_2=0.64$},
\mbox{$\beta_2=0.78$}, \mbox{$c_3=-0.88$}, \mbox{$\gamma_3=0.47$},
\mbox{$\beta_3=-0.88$}, \mbox{$c_4=-0.36$}, \mbox{$\gamma_4=0.78$},
\mbox{$\beta_4=-0.38$}, \mbox{$c_6=0.70$}, \mbox{$\delta_6=-0.89$},
\mbox{$\gamma_6=0.75$}, \mbox{$\beta_6=-0.74$}, \mbox{$c_7=-0.95$},
\mbox{$\delta_7=-0.80$}, \mbox{$\gamma_7=0.41$},
\mbox{$\beta_7=0.95$}.  } , shown in Table~\ref{tab:pointE}.  We see
that this point has an unusual Higgs hierarchy, since here $h$ is
heavier than both $A$ and $H^{\pm}$ (in this respect similar to point
C).  The signal for $h \to WW $, normalized to the SM, is $0.54$.  The
CP-even $H$ will be discovered first and will appear to be the SM
Higgs, since the signal is very close to the SM one.  Soon after, $h$
will be found in both the $ZZ$ and $WW$ channels, thus providing a
clear evidence of new physics.  Notice that here the coupling of $h$
to gauge bosons is extremely small, but due to kinematics it still
decays preferentially into gauge bosons.  We stress again that this is
a unique characteristic of the BMSSM Higgs sector in the low $\tan
\beta$ regime, since in the MSSM a similar behavior can only occur for
$H$.  For this point, the $1/M^2$ operators contribute about $30~{\rm
GeV}$ to $m_{h}$.

\medskip
\noindent
\textit{Point F: Multi-Higgs decay chains}

\begingroup
\squeezetable
\begin{table}
\begin{center}
\textbf{ POINT F} 
\end{center}
\begin{center}
\begin{tabular}{|c|c|c|c|}
\hline
$m_A ~({\rm GeV})$ & $m_h ~({\rm GeV}) $ & $m_H ~({\rm GeV}) $ & $m_{H^{\pm}} ~({\rm GeV}) $ \\
\hline
64 & 135 & 155 & 125 \\
\hline
$g_{hWW}^2$ & $g_{HWW}^2$  &  $g_{hgg}^2$ & $g_{Hgg}^2$ \\
\hline
0.002 & 0.991 &  0.65 & 1.17 \\
\hline
channel & BMSSM & channel & BMSSM  \\ 
\hline
$h \to b \bar{b} $     & 0.15  & $ h \to AA$  & 0.84  \\
$ H \to WW    $	 	    & 0.12  &
$ H \to AA     $     	    & 0.84  \\
$H \to b \bar{b} $            & 0.02  &
$ A \to b \bar{b} $            & 0.92  \\
$ H^{+} \to \tau \nu_{\tau}$     & 0.56  &
$H^{\pm} \to  W^{\pm} + A $ 			& $0.40$    \\
\hline
\end{tabular}
\end{center}
\caption{ {\em Masses and branching fractions in the BMSSM for point F. }}
\label{tab:pointF}
\end{table}
\endgroup
For the points outside the Tevatron reach with $m_h$ below $140~{\rm
GeV}$, the most remarkable feature is the possibility of having the
channels $h \to AA$ and $H \to AA$ kinematically open.  For the points
where these channels are closed, the situation is not as interesting,
so we will focus on the first scenario.  As an example, we show point
$F$~\footnote{The Lagrangian parameters for this point are:
\mbox{$\omega_1=1$}, \mbox{$\alpha_1=-1$}, \mbox{$c_1=0.89$},
\mbox{$\gamma_1=-0.62$}, \mbox{$\beta_1=-0.44$}, \mbox{$c_2=-0.27$},
\mbox{$\gamma_2=-0.47$}, \mbox{$\beta_2=0.83$}, \mbox{$c_3=-0.21$},
\mbox{$\gamma_3=0.72$}, \mbox{$\beta_3=-0.75$}, \mbox{$c_4=-0.34$},
\mbox{$\gamma_4=0.85$}, \mbox{$\beta_4=-0.74$}, \mbox{$c_6=0.89$},
\mbox{$\delta_6=-0.50$}, \mbox{$\gamma_6=-0.43$},
\mbox{$\beta_6=-0.87$}, \mbox{$c_7=-0.08$}, \mbox{$\delta_7=-0.75$},
\mbox{$\gamma_7=0.76$}, \mbox{$\beta_7=0.52$}.  } in
Table~\ref{tab:pointF}, where these channels are the dominant decay
modes of both $h$ and $H$.  For this point, the $1/M^2$ operators
contribute about $10~{\rm GeV}$ to $m_{h}$.  Focusing on the $AA$
channel, the possible final states for $h$ and $H$ are $b \bar{b} b
\bar{b}$, $b \bar{b} \tau \bar{\tau}$ and $\tau \bar{\tau} \tau
\bar{\tau}$~\cite{Carena:2007jk,Dermisek:2008uu,Cheung:2007sva}.  The
first one is very challenging due to the enormous QCD background,
while the third one suffers from a reduced signal [$BR(A \to \tau
\bar{\tau}) \sim 10 \%$].  This leaves the $b \bar{b} \tau \bar{\tau}$
channel as the most promising one.  For the case of $H$, one may also
look at the $gg \to H \to WW$ channel, whose signal is 1/5 of the SM
value, and could be discovered with about 100
fb$^{-1}$~\cite{atlasphystdr}.  For the charged Higgs, the dominant
decay mode is $\tau \nu_{\tau }$.  Notice also that $H^{\pm}\to
AW^{\pm}$ can have a sizable branching fraction, offering the
possibility to discover both $A$ and $H^{\pm}$ simultaneously in this
decay mode.


\subsection{Large $\tan\beta$ searches: general features}
\label{sec:largetb}

In this subsection we present our analysis for the large $\tan \beta$
regime, fixing $\tan \beta=20$.  As shown in
Section~\ref{sec:spectra}, the changes in the spectrum with respect to
the MSSM are less important than in the low $\tan \beta$ case.

We use Eq.~(\ref{eq:ggheff}) to estimate the gluon fusion production
cross section at NLO in $\alpha_{s}$.  Although the impact of the
bottom loop in the $K$-factor is more important for larger
$\tan\beta$, the NLO K-factor in our model is still expected to be
within $20\%$ of the NLO SM K-factor, as discussed at the beginning of
Section~\ref{sec:lowtb}.  Furthermore, as shown in
\cite{Spira:1997dg}, the effects on the K factor due to a light
sparticle spectrum like the one we are considering are negligible at
large $\tan\beta$.  Hence, we conclude that simply computing the
right-hand side of Eq.~(\ref{eq:ggheff}) allows us to obtain the NLO
gluon fusion production cross section within $20\%$ accuracy even at
large $\tan\beta$.  In this regime, production in association with a
$b\bar{b}$ pair can become important, and can be obtained in our model
from existing results by a simple rescaling with the effective
coupling $g^{2}_{\phi bb}$, where $\phi = h, H, A$ [see
Eq.~(\ref{geff})].

\begin{figure*}[t]
\begin{center}
\includegraphics[width=7.9cm]{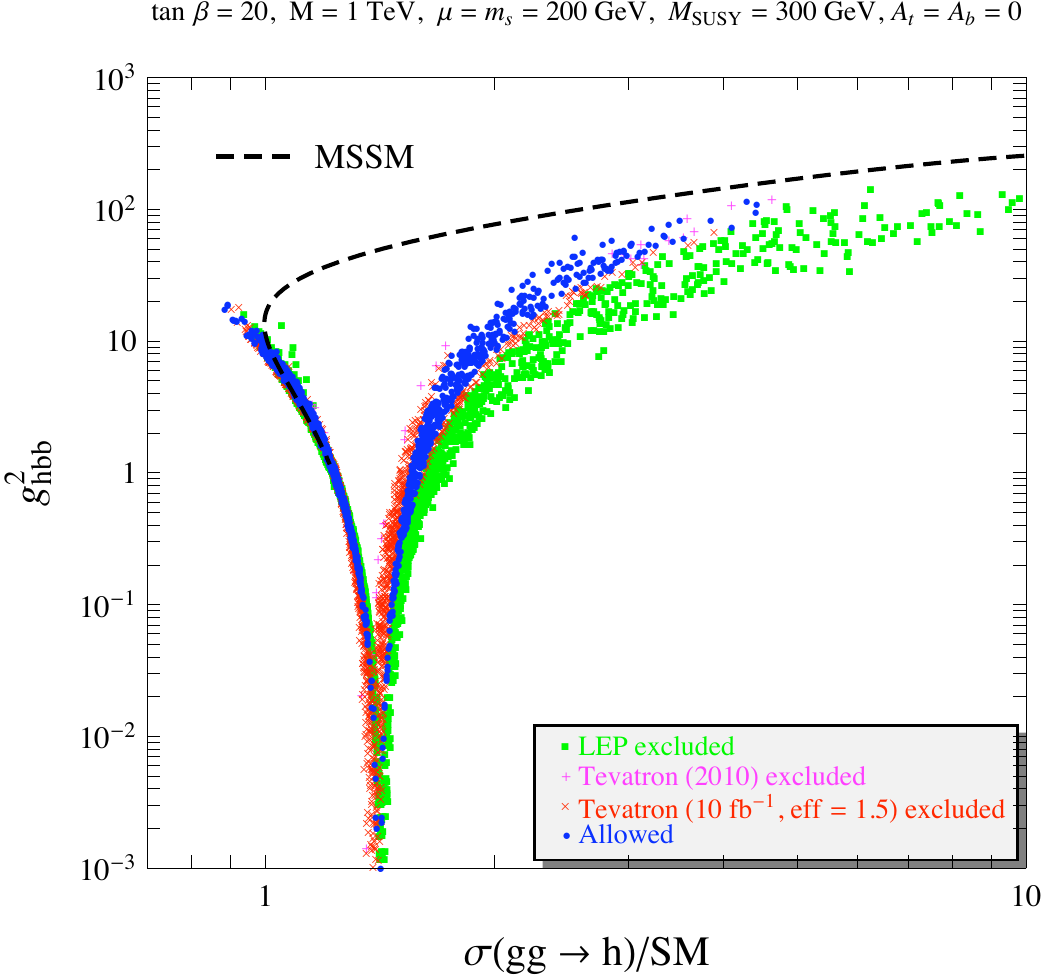}
\hspace{3mm}
\includegraphics[width=7.9cm]{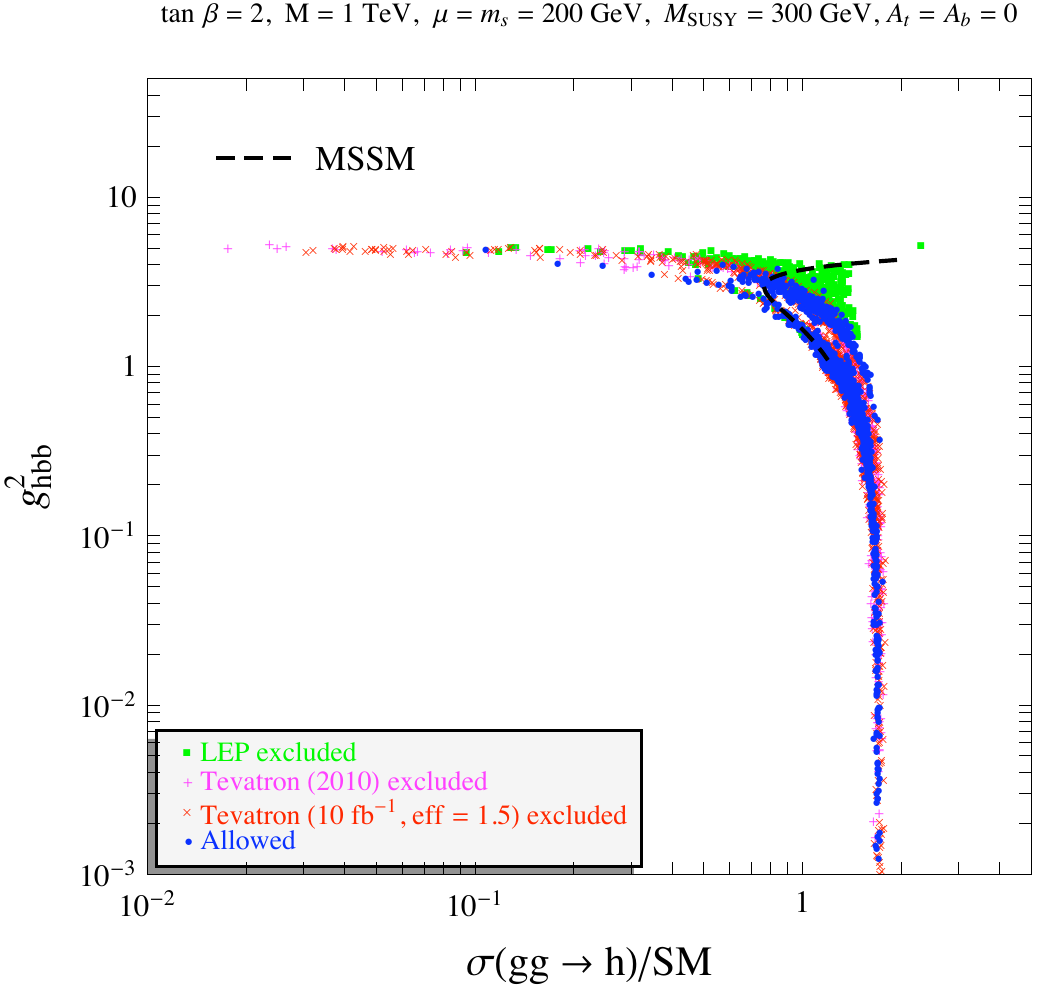}
\end{center}
\vspace{-0.5cm} 
\caption{\label{fig:lbblwwvsggl}{\em Effective
coupling $g^{2}_{h b \bar{b}}$ for $\tan\beta = 20$ (left panel), and
for $\tan\beta = 2$ (right panel), as a function of the gluon fusion
production cross section over the SM value.  We show the points
excluded by LEP (green), excluded by current Tevatron data (magenta)
and the region that will be probed by the Tevatron in the near future
(red).  The blue points are allowed by all the current and near future
experimental constraints.  The dashed line corresponds to the MSSM
result for the given SUSY spectrum.}}
\end{figure*}
We show in Fig.~\ref{fig:lbblwwvsggl} the effective coupling of $h$ to
down-type fermions, $g_{hb\bar{b}}^2$, for both large (left panel) and
small (right panel) $\tan\beta$.  At large $\tan\beta$, one sees that
the currently allowed models (blue and red points) have a gluon fusion
production cross section which ranges from $0.7 - 5$ times the SM
value.  The most striking feature is that the coupling to bottom pairs
can be strongly suppressed.  For large $\tan \beta$ this happens for a
value of $\sigma (gg \to h) /SM$ of around $1.4$.  These models have
$110~{\rm GeV} \lesssim m_{h} \lesssim 150~{\rm GeV}$, where the
decays into $b\bar{b}$ of the SM Higgs are important.  The suppression
in $g_{hb\bar{b}}^2$ can also be observed at low $\tan\beta$ (right
panel).  In this case, the associated values of $m_{h}$ are in the
somewhat higher range from $120~{\rm GeV}$ to $250~{\rm GeV}$, with
the strongest suppressions occurring for $m_{h} > 150~{\rm GeV}$.

The suppression in the coupling to down-type fermions is somewhat
reminiscent of the small $\alpha_{\rm eff}$
scenario~\cite{Carena:2002qg,Carena:2005ek}, but there are important
differences.  In the small $\alpha_{\rm eff}$ scenario the $g_{h b
\bar{b}}$ coupling is suppressed as a result of a cancellation between
the tree-level and one-loop contributions.  This can happen at large
$\tan\beta$, where the radiative effect is enhanced at the same time
that the tree-level contribution is somewhat suppressed, thus allowing
for a cancellation.  Besides large $\tan\beta$, sizable values of $\mu
A_t/M_{SUSY}^2$ are necessary, and the cancellation is found to happen
only for certain values of $m_{A}$ (below or of order $200~{\rm GeV}$)
that are highly correlated with $\tan\beta$~\cite{Carena:2002qg}.  In
contrast, the suppression we find occurs as a result of a cancellation
between the tree-level MSSM contribution and those due to the
higher-dimension operators (we have checked that the picture remains
unchanged by turning off all loop effects).  Most importantly, the
fact that the suppression occurs at tree-level implies that the
couplings to bottom and tau pairs are simultaneously (and strongly)
suppressed.  This does not tend to happen in the small $\alpha_{\rm
eff}$ scenario, since the radiative enhancements for bottoms and taus
happen in different regions of parameter space.  Also, in spite of the
large number of parameters, there is a clear correlation between the
$g_{hb\bar{b}}^2$ suppression and the $\sigma(gg \to h)$ enhancement.
The increase in the gluon fusion cross section is due to the
destructive interference of the bottom loop in the gluon fusion cross
section, and also to light SUSY particles running in the loop.

\begin{figure*}[tb]
\begin{center}
\includegraphics[width=7.9cm]{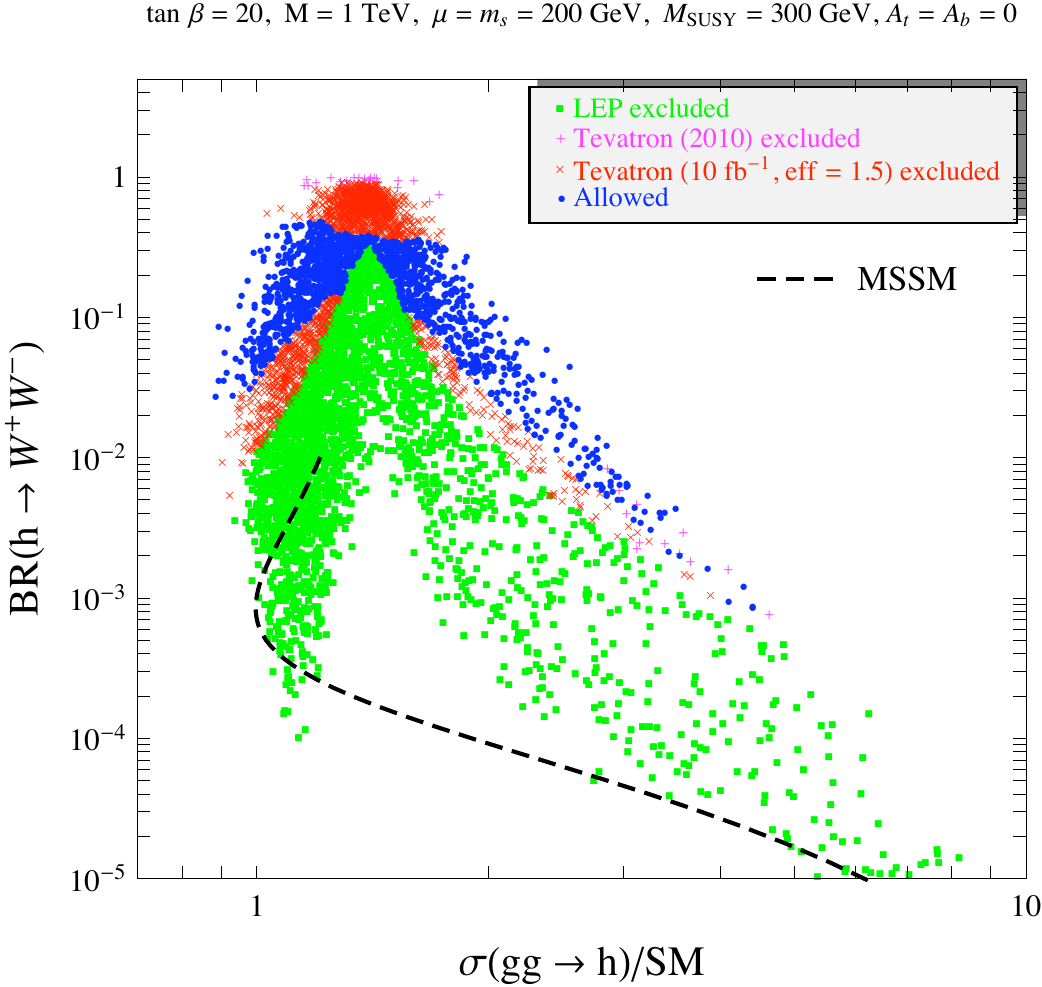}
\hspace{3mm}
\includegraphics[width =7.9cm]{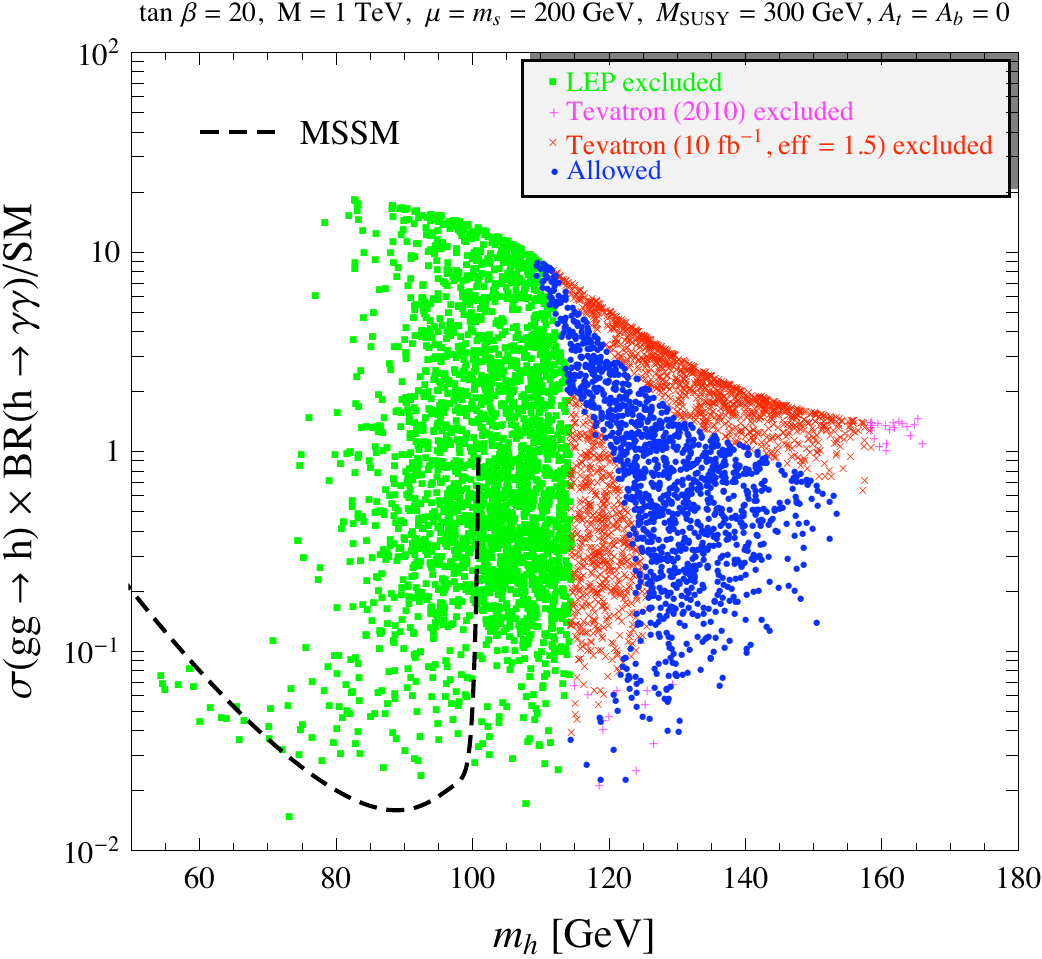}
\end{center}
\vspace{-0.5cm} 
\caption{\label{fig:htoggtb20}{\em Left panel: ${\rm
BR}(h \to W^+W^{-})$ as a function of the normalized gluon fusion
production cross section, for $\tan\beta = 20$.  Right panel: $\sigma
(gg \to h) \times BR(h \to \gamma \gamma)$, normalized to the SM
value, for $\tan\beta = 20$.  We show the points excluded by LEP
(green), excluded by current Tevatron data (magenta) and the region
that will be proved by the Tevatron in the near future (red).  The
blue points are allowed by all the current experimental constraints.
The dashed line corresponds to the MSSM result for the given SUSY
spectrum.}}
\end{figure*}
The suppression of the down-type fermion channels implies a general
enhancement in the branching fractions of the remaining channels.  The
most interesting enhancements are those in the gauge boson channels:
$WW$, $ZZ$ and $\gamma\gamma$.  In the left panel of
Fig.~\ref{fig:htoggtb20} we show the branching fraction into WW at
large $\tan\beta$, again as a function of the gluon fusion production
cross section normalized to the SM value, which can be compared to the
left panel of Fig.~\ref{fig:lbblwwvsggl}.  We see that the region
where the $hb\bar{b}$ coupling is suppressed is exactly where the WW
branching fraction is greatly enhanced, and leads to an interesting
Tevatron sensitivity in the $W$ channel over a wide range of $m_{h}$.
The left panel of Fig.~\ref{fig:htoggtb20} clearly exhibits how the
Tevatron covered (red) points arise.  The upper red region corresponds
to those models within Tevatron reach in the $h \to WW$ search, while
the two lower red regions contain only points that can be probed in
the $h \to b \bar{b}$ channel.  These latter models always have a
branching fraction into WW below about 20\%.

In the right panel of Fig.~\ref{fig:htoggtb20} we show the $gg \to h
\to \gamma \gamma$ cross section, normalized to the SM. We see that
the di-photon signal can be increased with respect to the SM one by up
to a factor of 10.  This strong enhancement is a direct result of the
decreased branching fraction into $b \bar{b}$, together with the
enhancement in the gluon production cross section discussed above.
The points with enhanced signal in the diphoton channel correspond to
values of $m_h$ between $110~{\rm GeV}$ and $130~{\rm GeV}$.  It is
interesting to compare to the latest available diphoton analysis from
CDF \cite{:cdftwogammas} and D0 \cite{:d0twogammas}.  The CDF
analysis, performed with 5.4 fb$^{-1} $ of data, quotes an observed
limit of 18.7-25.9 for the di-photon cross section normalized to the
SM. The D0 analysis, with 4.2 fb$^{-1} $, gives a corresponding limit
of 11.9-28.3 .~\footnote{The factor of almost ten enhancement in the
di-photon signal in our model is based on gluon fusion production,
which at the Tevatron contributes 73 \% - 95 \% of the total SM cross
section.} As a result, the enhancement in the di-photon signal we find
can be interesting at the Tevatron, and of course it would be
spectacular at the LHC. One should also notice that for models with
enhanced $BR(h \to b \bar{b})$, the signal into photons can be reduced
by up to a factor of 10.
 
The $gg \to h \to WW$ signal (not shown here) presents the same
behavior as the $\gamma \gamma$ one.  This can be easily understood as
follows.  In the SM, the $h \to \gamma \gamma$ decay mode proceeds via
$W$ and top loops, the former giving the dominant effect.  In our
currently allowed (blue and red) points, the coupling of $h$ to tops
and W's is very close to the SM value (the differences are below 2\%).
Although $g_{hb\bar{b}}$ can be enhanced by a factor of $10$, the bottom
loop is still a small contribution to the $h \to \gamma \gamma$
process.  Therefore, the partial widths $\Gamma (h \to WW)$ and
$\Gamma (h \to \gamma \gamma)$ in our model are very close to the SM
ones, and the changes in the branching ratios of each channel are
common and strictly due to the variation of $BR(h \to b \bar{b})$ with
respect to the SM. Therefore, enhancements in the $WW/ZZ$ channels can
also be interestingly large.

With respect to the remaining Higgs bosons, the situation resembles
the large $\tan \beta$ regime of the MSSM. Both $H$ and $A$ decay
mainly into bottoms and taus, while the charged Higgs goes to either
$\tau \nu_{\tau}$ or $t \bar{b}$ depending on its mass.  It is also
possible for a heavy Higgs to decay into the lightest one: ${\rm BR}(H
\to h h)$ can reach 30\%, while both ${\rm BR}(A \to h Z)$ and ${\rm
BR}(H^{\pm} \to h W^{\pm})$ can reach 10\%, provided the decaying
Higgs boson mass is above $200~{\rm GeV}$.  For this mass range, the
decay mode into sparticles can also be important, if kinematically
allowed.

\subsection{Large $\tan\beta$ searches: benchmark points}
\label{sec:largetbbench}

Having described the main differences of the large tangent beta regime
with respect to the MSSM, we show a selected sample of benchmark
points.

\subsubsection{Scenarios within Tevatron reach}

We start with the points covered in the near future by the Tevatron
via the $h \to b \bar{b}$ search.  From Fig.~\ref{fig:htoggtb20} one
sees that for these points (lowest red region in the left panel), the
signal into $\gamma \gamma$ (and $WW$) can be enhanced by at most a
factor of $2$.  Since such enhancement factors can also be obtained
within the MSSM (for instance with sparticle masses around $500~{\rm
GeV}$), we will not show a benchmark point here, but will
briefly comment on the main characteristics of these type of models.
The Tevatron could claim a hint in the $b \bar{b}$ channel, while at
the LHC the signals into $\gamma \gamma$ and $\tau \bar{\tau}$ are
enhanced by up to a factor of $2$ with respect to the SM, thus
allowing for a discovery using these decays modes.  Regarding the
remaining Higgs bosons, one sees that both $H$ and $A$ decay mainly
into bottoms and taus.  The gluon fusion production cross section for
$H$ and $A$ is around 80\% of the SM-value, while the $bbh$ production
becomes an important mechanism due to the large $\tan\beta$
enhancement (the CP-even $H$ has highly suppressed couplings to W's
and Z's).  Thus a discovery in the $H/A \, \tau\bar{\tau}$ search may
be feasible~\cite{atlasphystdr}.  We note also that $A$ and $H$ can be
very close in mass, so that the two states cannot be disentangled at
the LHC, but rather the signals have to be added up.  Because of a sizable
branching fraction into the $\tau \nu_{\tau}$ channel, the charged
Higgs can be within LHC reach, even for $m_{H^{\pm}} > m_{t}$
\cite{Aad:2009wy}.  The way to distinguish such a situation from the
MSSM will be through the observation of relatively light
superparticles.

As we mentioned before, an interesting possibility is to have sizable
branching fractions for the decay modes $H \to h h$, $A \to h Z$ and
$H^{\pm} \to h W^{\pm}$.  This requires heavy Higgs bosons with a mass
above $250 ~{\rm GeV}$.  Depending on the details of the SUSY
spectrum, also decays into sparticles may be open.  We have found that
the branching fractions in these multi-Higgs channels are below 10\%
in most cases, and that one would still have both $A$ and $H$ decaying
sizably into down-type fermions, with $H^{+}$ decaying preferably into
$t \bar{b}$ but with a non-negligible branching fraction into
$\tau\nu_{\tau}$ due to the large $\tan\beta$ enhancement.  Provided
that $m_{H} > 300~{\rm GeV}$, the $H \to h h$ branching fraction can
reach values of up to $20 - 30\%$, which is interesting since it allows for
the potential observation of several Higgs states.

\medskip
\noindent
\textit{ Point G: SM-like Higgs heavier than the MSSM upper bound}

\begingroup
\squeezetable
\begin{table}
\begin{center}
\textbf{ POINT G} 
\end{center}
\begin{center}
\begin{tabular}{|c|c|c|c|}
\hline
$m_A ~({\rm GeV})$ & $m_h ~({\rm GeV}) $ & $m_H ~({\rm GeV}) $ & $m_{H^{\pm}} ~({\rm GeV}) $ \\
\hline
267 & 148.6 & 297 & 283 \\
\hline
$g_{hWW}^2$ & $g_{HWW}^2$  &  $g_{hgg}^2$ & $g_{Hgg}^2$ \\
\hline
0.97 & 0.03 & 1.64 & 0.14 \\
\hline
channel & BMSSM (SM) & channel & BMSSM (SM)  \\ 
\hline
$h \to b \bar{b} $     & 0.43 (0.20) & $ h \to \tau \bar{\tau}  $ & 0.07 (0.02)  \\
$h \to ZZ $     & 0.08 (0.05) & $ h \to WW $  & 0.41 (0.66)  \\
$ H \to b \bar{b}    $	 	    & 0.75  &
$ H \to \tau \bar{\tau}     $     	    & 0.13  \\
$A \to b \bar{b} $            & 0.84  &
$ A \to \tau \bar{\tau} $            & 0.14  \\
$ H^{\pm} \to \tau \nu_{\tau} $     & 0.21  &
$H^{\pm} \to  t \bar{b} $ 			& 0.75  \\
\hline
\end{tabular}
\end{center}
\caption{ {\em Masses and branching fractions in the BMSSM (and in the
SM for $h$) for point G.  }}
\label{tab:pointG}
\end{table}
\endgroup
We turn now to the models covered at the Tevatron via the $h \to WW$
search (upper red region in Figs.~\ref{fig:htoggtb20}).  We show in
Table~\ref{tab:pointG}~\footnote{The Lagrangian parameters for this
point are: \mbox{$\omega_1=1$}, \mbox{$\alpha_1=-1$}, \mbox{$c_1=0$},
\mbox{$\gamma_1=0.80$}, \mbox{$\beta_1=-0.63$}, \mbox{$c_2=-0.08$},
\mbox{$\gamma_2=0.69$}, \mbox{$\beta_2=0.52$}, \mbox{$c_3=-1$},
\mbox{$\gamma_3=-0.74$}, \mbox{$\beta_3=0.44$}, \mbox{$c_4=-0.73$},
\mbox{$\gamma_4=0.41$}, \mbox{$\beta_4=-0.90$}, \mbox{$c_6=0.64$},
\mbox{$\delta_6=-0.62$}, \mbox{$\gamma_6=0.40$},
\mbox{$\beta_6=-0.66$}, \mbox{$c_7=-0.56$}, \mbox{$\delta_7=-0.52$},
\mbox{$\gamma_7=-0.66$}, \mbox{$\beta_7=-0.71$}.  } an example where
$m_h$ is above the maximum attainable value in the $m_h$ max scenario
of the MSSM, with sparticles at the ${\rm TeV}$ scale.  The $1/M^2$
operators contribute about $45~{\rm GeV}$ to $m_{h}$.  We notice that
in this point the enhancement in the gluon fusion production cross
section [$1.64 \times \sigma_{SM}(gg \to h)$] is compensated by the
reduction in the $WW$ branching fraction with respect to the SM
($0.41/0.66$), thus resulting in a $gg \to h \to WW$ signal close to
the SM one.  As can be seen in the right panel of
Fig.~\ref{fig:htoggtb20} (interpreted for the $WW$ channel) this is a
general feature of the Tevatron covered points in the higher range of
$m_{h}$.  The nonstandard neutral Higgs bosons, $H$ and $A$, can be
detected in the $\tau \bar{\tau}$ (or $\mu \bar{\mu}$) channels, as is
well known for the large $\tan\beta$ region of the MSSM. The charged
Higgs can be searched for in the $\tau\nu_{\tau}$ channel.  However,
we emphasize again that the observation of light SUSY signals would
give compelling evidence for BMSSM physics.
There are also Tevatron covered (red) points at smaller $m_{h}$
values, around $110~{\rm GeV}$ with an enhanced di-photon signal.  We
discuss these type of scenarios in the next section, together with the
Tevatron uncovered (blue points) in the same region.

\subsection{LHC searches}

Referring to Fig.~\ref{fig:htoggtb20} we split the Tevatron uncovered
(blue) points according to whether their signal into photons (and W's)
is enhanced or suppressed.  For the latter case, one has that $h$
decays mainly into bottom and tau pairs.  In these scenarios, $h$ can
be within the reach of the LHC in the $\tau \bar{\tau}$ channel, and
if the suppression of the ZZ coupling is not extreme, maybe also in
the $ZZ \to 4l$ channel.  Higgs decay chains, such as $H \to hh \to b
\bar{b} \tau \bar{\tau}$, can also give rise to interesting (if
challenging) signatures.  We do not show a benchmark point here since
the branching fractions of the relevant Higgs decay chain modes will 
depend on the details of the sparticle spectrum.

\medskip
\noindent
\textit{ Point H: SM-like Higgs with enhanced di-photon signal}

\begingroup
\squeezetable
\begin{table}
\begin{center}
\textbf{ POINT H} 
\end{center}
\begin{center}
\begin{tabular}{|c|c|c|c|}
\hline
$m_A ~({\rm GeV})$ & $m_h ~({\rm GeV}) $ & $m_H ~({\rm GeV}) $ & $m_{H^{\pm}} ~({\rm GeV}) $ \\
\hline
210 & 111.3 & 215 & 225 \\
\hline
$g_{hWW}^2$ & $g_{HWW}^2$  &  $g_{hgg}^2$ & $g_{Hgg}^2$ \\
\hline
0.98 & 0.02 & 1.39 & 0.84 \\
\hline
channel & BMSSM (SM) & channel & BMSSM (SM)  \\ 
\hline
$h \to b \bar{b} $     & 0.03 (0.79) & $ h \to \gamma \gamma / 10^{-3}$  & 12.1 (2.1)  \\
$h \to $ jets     & 0.56 (0.07) & $ h \to WW $  & 0.36 (0.05)  \\
$ H \to b \bar{b}    $	 	    & 0.86  &
$ H \to \tau \bar{\tau}     $     	    & 0.14  \\
$A \to b \bar{b} $            & 0.86  &
$ A \to \tau \bar{\tau} $            & 0.14  \\
$ H^{\pm} \to \tau \nu_{\tau} $     & 0.35  &
$H^{\pm} \to  t \bar{b} $ 			& 0.64  \\
\hline
\end{tabular}
\end{center}
\caption{ {\em Masses and branching fractions in the BMSSM (and in the
SM for $h$) for point H. }}
\label{tab:pointH}
\end{table}
\endgroup
We illustrate the features of models with a strong enhancement of the
di-photon signal with point H~\footnote{The Lagrangian parameters for
this point are: \mbox{$\omega_1=0.50$}, \mbox{$\alpha_1=-0.93$},
\mbox{$c_1=0.88$}, \mbox{$\gamma_1=-0.41$}, \mbox{$\beta_1=-0.76$},
\mbox{$c_2=0.07$}, \mbox{$\gamma_2=-0.67$}, \mbox{$\beta_2=-0.63$},
\mbox{$c_3=-0.60$}, \mbox{$\gamma_3=0.55$}, \mbox{$\beta_3=0.74$},
\mbox{$c_4=-0.61$}, \mbox{$\gamma_4=-0.58$}, \mbox{$\beta_4=0.68$},
\mbox{$c_6=0.10$}, \mbox{$\delta_6=-0.99$}, \mbox{$\gamma_6=-0.45$},
\mbox{$\beta_6=-0.37$}, \mbox{$c_7=0.59$}, \mbox{$\delta_7=-0.75$},
\mbox{$\gamma_7=-0.93$}, \mbox{$\beta_7=0.99$}.  } (shown in
Table~\ref{tab:pointH}).  We see that $h$ is rather light (the $1/M^2$
operators contribute about $10~{\rm GeV}$ to $m_{h}$), but escaped
detection at LEP due to the strong suppression of the $b\bar{b}$
channel.  The $gg \to h \to \gamma \gamma$ signal is larger than the
SM one by a factor of $8$, thus allowing for a very nice and clean
detection of $h$ at the LHC. As was discussed in the context of
Fig.~\ref{fig:htoggtb20}, the same enhacement also occurs for the $WW$
and $ZZ$ channels.  Therefore, and in spite of such a light Higgs
mass, the $gg \to h \to ZZ \to 4l$ channel would be at the reach of
the LHC.

For the remaining neutral Higgs bosons ($H$ and $A$), one will have to
consider the $\tau \bar{\tau}$ search.  The charged Higgs may be
detected at the LHC in the $\tau \nu_{\tau}$ channel.

Note that the benchmark point H has nonstandard Higgs bosons that are too
light to allow decays into $hh$.  However, given that $h$ is rather
light in the region with suppressed $b\bar{b}$ couplings, it is
possible that such \emph{exotic} channels might be open, while still
having an interesting di-photon signal.  As mentioned before, in such
cases it is possible that other channels involving SUSY particles are
also open.

\section{Conclusions}
\label{sec:conclu}

We studied the Higgs collider phenomenology of BMSSM scenarios, i.e.
supersymmetric extensions of the MSSM within an EFT framework where
the effects of the BMSSM degrees of freedom enter through
higher-dimension operators.  As emphasized in \cite{Carena:2009gx} the
first two orders in the $1/M$ expansion can be phenomenologically
significant, and should be included.  In the present work, we have
performed a model-independent study to highlight the variety of
collider signals that become available in such scenarios.

The coupling of the lightest CP-even Higgs to bottom pairs can be
suppressed due to cancellations between the MSSM contribution and
those from the higher-dimension operators.  It does not seem to
require a special tuning of parameters and occurs in both the low and
large $\tan\beta$ regimes.  As a result, the signals in clean
channels, such as the di-photon or $WW$ ones, can be greatly enhanced.
This suppression in the $h b \bar{b}$ induces an enhancement in the
gluon fusion production cross section, beyond the one arising from
light sparticles in the loop.

To emphasize the interplay between the Tevatron and the LHC, we have
analyzed projections for the Tevatron assuming a total integrated
luminosity of $10~{\rm fb}^{-1}$ per experiment and a 50\% efficiency
improvement in the $WW$ and $b\bar{b}$ search channels with respect to
present results.  We find that the current Tevatron data already
probes a large class of SUSY models, especially in the $WW$ channel.
The future projections indicate that the $b \bar{b}$ channel can
become effective for a SM-like Higgs search.  Moreover, a combination
of the $b\bar{b}$ and $WW$ search channels, together with the
$\tau\bar{\tau}$ decay mode in the large $\tan \beta$ region, would
further enlarge the set of BMSSM models that can be probed at the
Tevatron.  However, our main interest in this work was to survey the
types of signals that might be expected in SUSY scenarios, many of
which are not realized in the MSSM limit.  Improving the analysis by
combining channels and/or moderately increasing the luminosity will
not significantly change our conclusions.  Lightest CP-even Higgs
bosons with masses above $180~{\rm GeV}$, that can not be probed by the 
Tevatron, will be at the reach of the LHC.

Most of the changes in the expected Higgs signals, compared to the
MSSM, can be understood in large part from the altered Higgs spectrum.
We have surveyed a wide range of possibilities by scanning over the
parameter space of the higher-dimension operators.  Motivated by
naturalness arguments, we have chosen the SUSY-breaking scale close to
the EW scale, with the BMSSM physics at the ${\rm TeV}$ scale.  In
this case, the contributions from the SUSY particles to the Higgs
spectrum are subleading compared to the ones coming from the BMSSM
physics.  In the case of a heavier SUSY spectrum, and for a scale $M$
such that the effective field theory approach remains valid, the
qualitative features of the Higgs phenomenology triggered by the BMSSM
physics will be similar.  However, a detailed study should be
performed for each specific choice of the heavy scale $M$, the scale
of SUSY-breaking $m_S$, and the $\mu$-term to address the quantitative
features of the Higgs sector.

We have defined a number of ``benchmark points'' in order to discuss
the correlations between different Higgs signals.  Interestingly, we
find that there can be significant mixing in the CP-even Higgs sector,
allowing non-negligible couplings of both CP-even Higgs eigenstates to
the EW gauge bosons.  In addition, they can both be in the right mass
range to decay predominantly into $W$'s or $Z$'s, thus enabling a
detailed and direct study of the physics of EWSB. For these benchmark
points, the $1/M^2$ effects add a few tens of GeV to $m_{h}$, and have
a rather relevant impact on the collider phenomenology (but we remind
the reader that the $1/M^2$ operators can easily give a much larger
contribution to $m_{h}$; see Fig.~1 of Ref.~\cite{Carena:2009gx}).
Furthermore, we have found viable examples where the nonstandard
CP-odd Higgs can be produced in charged Higgs decays.  Moreover,
unusual decay chains such as $h \to AA$ or $H \to AA$ are also
possible, without $A$ being ultralight.  These channels are most
interesting in the low $\tan\beta$ region where the
$\tan\beta$-enhanced production of the nonstandard Higgs bosons is
not available.  These Higgs decay chains open the possibility of fully
reconstructing the Higgs content of a 2HDM in such supersymmetric
scenarios.  We also find scenarios where observing the Higgs sector is
more challenging, and would require dedicated studies that go beyond
the scope of this work.

 In conclusion, we find that Higgs signals in supersymmetric scenarios
 can be markedly different from those in the MSSM paradigm.  If all
 third generation squarks turn out to be light ($m_S \leq 300~{\rm
 GeV}$), given the LEP Higgs mass bounds, this will imply a clear case
 for BMSSM physics.  The heavier degrees of freedom could be at the
 kinematic reach of the LHC, but depending on their nature the direct
 discovery might be elusive.  In either case, supersymmetric Higgs
 searches can provide evidence of physics beyond the MSSM.

\begin{acknowledgements}
We would like to thank Oliver Brein and Karina Williams for making an
unofficial version of the HiggsBounds code available to us, and for
the help provided.  J.Z would like to thank the Theory Division of
Fermilab for hospitality during the final stages of this work.
Fermilab is operated by the Fermi Research Alliance, LLC under Contract
No.  DE-AC02-07CH11359 with the U.S. Department of Energy.  E.P. is
supported by DOE Grant No. DE-FG02-92ER40699.  The work of J.Z is
supported by the Swiss National Science Foundation (SNF) under
Contract No. 200020-126691.
\end{acknowledgements}



\end{document}